\documentclass{emulateapj}

\usepackage{aas_macros}
\usepackage{amssymb}
\usepackage{txfonts}
\usepackage{natbib}
\usepackage[svgnames]{xcolor}
\makeatletter

\newcommand{\Rmnum}[1]{\expandafter\@slowromancap\romannumeral #1@}
\makeatother
\newcommand  \NV{\,N\,{\footnotesize V} }

\newcommand  \HII{\,H\,{\footnotesize II} }
\newcommand  \HI{\,H\,{\footnotesize I} }
\newcommand  \m{\mathrm}

\begin{document}

\title{Nested shells reveal the rejuvenation of the Orion-Eridanus superbubble}

\author{Bram B. Ochsendorf\altaffilmark{1}, Anthony G. A. Brown\altaffilmark{1}, John Bally\altaffilmark{2} \& Alexander G. G. M. Tielens\altaffilmark{1}}
\affil{$^1$ Leiden Observatory, Leiden University, P.O. Box 9513, NL-2300 RA, The Netherlands\\ $^2$ CASA, University of Colorado, CB 389, Boulder, CO 80309, USA}
\email{ochsendorf@strw.leidenuniv.nl}

\begin{abstract}
The Orion-Eridanus superbubble is the prototypical superbubble due to its proximity and evolutionary state. Here, we provide a synthesis of recent observational data from {\em WISE} and {\em Planck} with archival data, allowing to draw a new and more complete picture on the history and evolution of the Orion-Eridanus region. We discuss the general morphological structures and observational characteristics of the superbubble, and derive quantitative properties of the gas- and dust inside Barnard's Loop. We reveal that Barnard's Loop is a complete bubble structure which, together with the $\lambda$ Ori region and other smaller-scale bubbles, expands within the Orion-Eridanus superbubble. We argue that the Orion-Eridanus superbubble is larger and more complex than previously thought, and that it can be viewed as a series of nested shells, superimposed along the line of sight. During the lifetime of the superbubble, \HII region champagne flows and thermal evaporation of embedded clouds continuously mass-load the superbubble interior, while winds or supernovae from the Orion OB association rejuvenate the superbubble by sweeping up the material from the interior cavities in an episodic fashion, possibly triggering the formation of new stars that form shells of their own. The steady supply of material into the superbubble cavity implies that dust processing from interior supernova remnants is more efficient than previously thought. The cycle of mass-loading, interior cleansing, and star formation repeats until the molecular reservoir is depleted or the clouds have been disrupted. While the nested shells come and go, the superbubble remains for tens of millions of years.
\end{abstract}

\section{Introduction}

The Orion-Eridanus superbubble is a nearby ($\sim$400 pc) expanding structure that is thought to span 20$^{\circ}$ $\times$ 45$^{\circ}$ on the sky \citep{reynolds_1979,bally_2008,pon_2014b}. Because of its proximity and evolutionary stage, it is traced at a multitude of wavelengths \citep{heiles_1999}, serving as a benchmark to the study of superbubbles. The expansion of the bubble is most likely connected to the combined effects of ionizing UV radiation, stellar winds, and a sequence of supernova explosions from the Orion OB association, Orion OB1 \citep{blaauw_1964,brown_1994}.

The line of sight expansion of the Orion-Eridanus superbubble is previously estimated at $\sim$ 15 km s$^{-1}$ through line-splitting of H$\alpha$, which also revealed a total ionized gas mass of 8 $\times$ 10$^{4}$ M$_{\odot}$ and a kinetic energy of $E_\m{kin}$ \textgreater\ 1.9 $\times$ 10$^{50}$ erg \citep{reynolds_1979}. Large-scale spectroscopic mapping of the entire Orion-Eridanus region in \HI reported a larger expansion velocity of the superbubble of $\sim$ 40 km s$^{-1}$ with a mass of 2.5 $\times$ 10$^5$ M$_\m{\odot}$, containing a kinetic energy of 3.7 $\times$ 10$^{51}$ erg \citep{brown_1995}, which is shown to be consistent with the integrated mechanical luminosity exerted by Orion OB1 ($\sim$ 10$^{52}$ erg; \citealt{brown_1995}). The expansion of the superbubble is also traced through high-velocity and intermediate-velocity gas in several lines of sight towards Orion \citep{cowie_1979,welty_2002}. Soft X-rays emanate from the million-degree plasma in the interior of the superbubble \citep{burrows_1993,snowden_1995}.

Here, we present observations of recent all-sky surveys with {\em WISE} and {\em Planck}, and combine them with existing sky surveys to provide a new and more complete insight in the history and evolution of the Orion-Eridanus region. In particular, we will reveal that Barnard's Loop is part of a complete bubble structure, separate from the Orion-Eridanus superbubble, that sweeps up the mass-loaded interior of the pre-existing superbubble. It is argued that the Orion-Eridanus superbubble is larger and more complex than previously thought, and that the entire morphological appearance of the superbubble can be viewed as a series of nested shells, superimposed along the line of sight. The shells originate from explosive feedback from Orion OB1 that accelerate, sweep-up, and compress the superbubble interior plasmas in an episodic fashion to form nested shells within the Orion-Eridanus superbubble. We explore the origin of the shells, their relation with the subgroups of Orion OB1, and their impact on the molecular clouds and star formation efficiency within Orion. We discuss our findings in terms of the long-term evolution of the superbubble. We present the obervations in Sec. \ref{sec:obs} and our results in Sec. \ref{sec:results}. In Sec. \ref{sec:discussion}, we discuss our findings, and summarize in Sec. \ref{sec:summary}.

\section{Observations}\label{sec:obs}

We made use of the all-sky surveys from {\em Planck} \citep{planck_collaboration1_2014}, the {\em Wide-Field Infrared Explorer} (WISE; \citealt{wright_2010}), the Leiden/Argentina/Bonn survey of Galactic \HI (LAB; \citealt{kalberla_2005}), the {\em R\"{o}ntgensatellit} soft X-ray background (ROSAT SXRB; \citealt{snowden_1997}), and an all-sky H$\alpha$ map that combines several large-scale surveys \citep{finkbeiner_2003}, including the {\em Southern H-alpha Sky Survey Atlas} (SHASSA; \citealt{gaustadt_2001}), the {\em Virginia-Tech Spectral-line Survey} (VTSS; \citealt{dennison_1998}), and the {\em Wisconsin H$\alpha$ Mapper} (WHAM; \citealt{haffner_2003}).

\section{Results}\label{sec:results}

\begin{figure*}
\centering
\includegraphics[width=17cm]{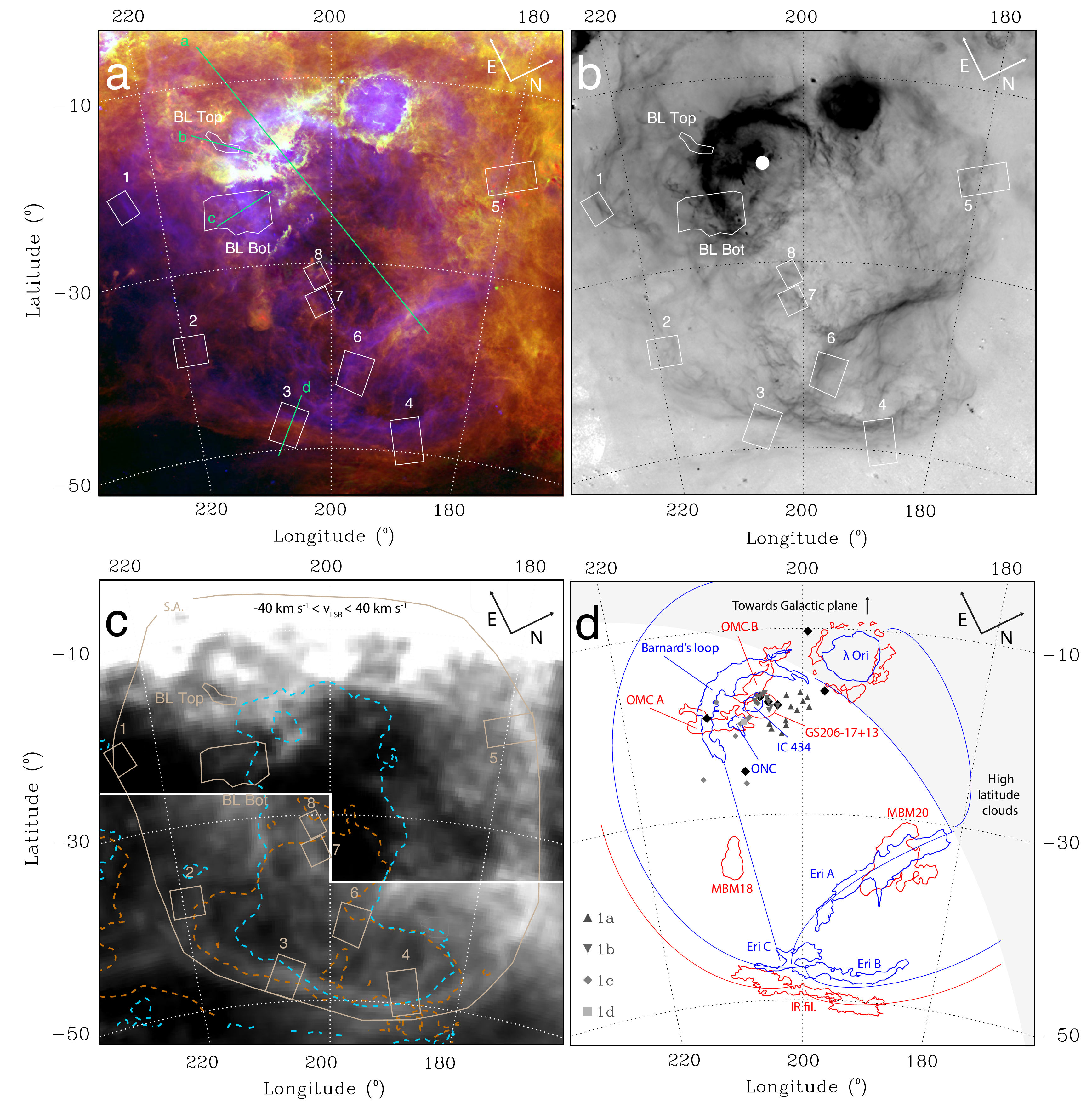} 
\caption{{\bf (a)} Three-color image of the Orion-Eridanus superbubble. H$\alpha$ (blue) reveals ionized gas, surrounded by a layer of polycyclic aromatic hydrocarbons or stochastically-heated small grains, traced by the WISE 12 $\mu$m band (green). Red is Planck 353 GHz that probes columns of cold (20 K) dust. The polycyclic aromatic hydrocarbons/small grains trace the UV-illuminated edges of the clouds containing cold dust, resulting in that red and green blend to form yellow. To the north-east (top part the image) lies the Galactic plane, while to the north-west (right part of the image) lie high-latitude clouds \citep[e.g.,][]{schlafly_2014a}. The numbered boxes define the apertures discussed in Sec. \ref{sec:apertures}, while the light-green labelled solid lines mark the positions of cross-cuts used in Fig. \ref{fig:cuts}. {\bf (b)} Same region as panel (a), but only showing the H$\alpha$ emission using a different stretch to accentuate the faint shell structure which envelops the Orion region in H$\alpha$. The white dot reveals the flux-weighted centre of the H$\alpha$ bubble \citep{reynolds_1979}. {\bf (c)} HI emission from the LAB survey, integrated in velocity space along the full extent of the Orion-Eridanus superbubble, -40 km s$^{-1}$ $\textless$ $v_\m{LSR}$ $\textless$ 40 km s$^{-1}$ \citep{brown_1995}. Overlayed are contours of ROSAT 0.25 keV (orange dashed) and ROSAT 0.75 keV diffuse X-rays (light blue dashed). The 0.75 keV emission projected within Barnard's Loop originates from bright sources within the ONC and IC 434 regions, but is blended with the more diffuse emission from the superbubble because of smoothing of the ROSAT maps. Below the white solid line the image has a different scaling to accentuate the faint HI filament intersecting apertures 2-4. Also overplotted is the shape of the superbubble aperture (S.A.; see text) {\bf (d)} Schematic representation of the region. H$\alpha$ gas structures are shown in blue lines/contours, while dust structures are plotted in red. The solid lines trace faint filaments of the region to guide the eye in panels (a) \& (b). The stars that form the well-known constellation of Orion are also plotted to appreciate the immense size of the region (black diamonds). Members of the different subgroups of the Orion OB association  \citep{blaauw_1964,brown_1994} with spectral type B2 or earlier are also shown in different grey symbols shown by the legend. See text for the explanation of the different abbreviations.} 
\label{fig:superbubble}
\end{figure*}

Figure 1 reveals the large-scale structure of the Orion-Eridanus superbubble. In H$\alpha$, the region exhibits various filamentary structures, including Barnard's loop \citep{barnard_1894} and the Eridanus filaments \citep[e.g.,][]{pon_2014a}. Projected on top of Barnard's Loop are the Orion A and Orion B molecular clouds (OMC A and OMC B), the Orion Nebular cluster (ONC) associated with the Orion nebula, and the IC 434 emission nebula that is characterized by a champagne flow of ionized gas (\citealt{tenorio_tagle_1979}; Sec. \ref{sec:smallershells}). Another prominent feature is the $\lambda$ Orionis bubble, a 10$^{\circ}$ circular \HII\ region, encapsulated by a swept-up shell of gas and dust. Below, we will refer to different parts of the superbubble in RA/Dec space (see the orientation in Fig. \ref{fig:superbubble}). 

\begin{figure*}
\includegraphics[width=18cm]{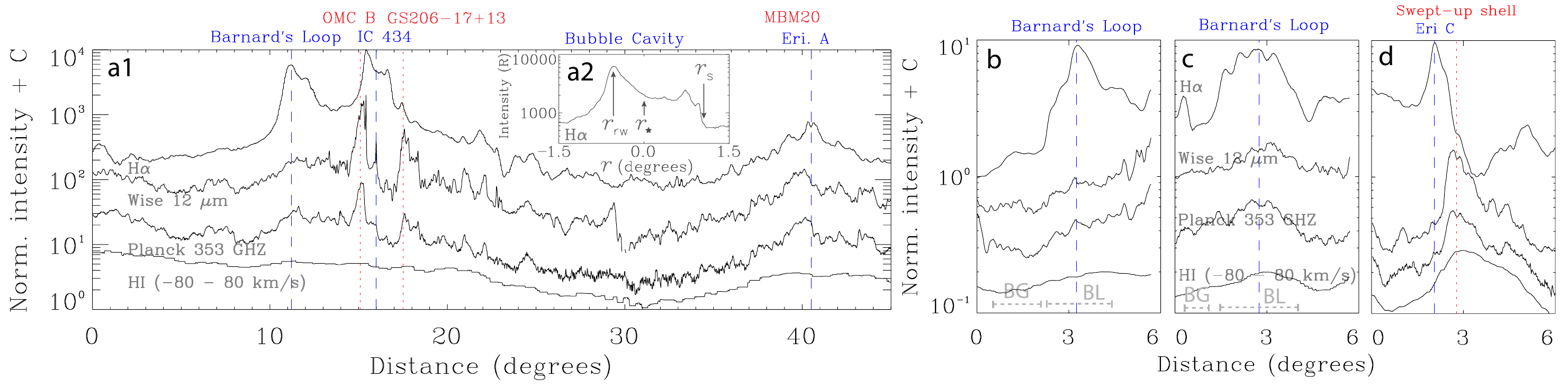} 
\caption{Profiles of cross-cuts defined in Fig \ref{fig:superbubble}a. Cuts start at the labelled side in Fig. \ref{fig:superbubble}a. {\bf (a1)} Along the cut from left to right: Barnard's loop; the Orion B molecular cloud (OMC B) and the IC 434 emission nebula, the GS206-17+13 shell; the excavated bubble interior; and the MBM 20 high-latitude cloud, overlapping the Eri A H$\alpha$ filament. The dashed vertical lines highlight several prominent features seen in the gas (blue) and in dust (red). {\bf (a2)} Inset of the IC 434 emission nebula H$\alpha$ profile, plotted as intensity (in units of Rayleigh) as a function of distance from the ionizing star $r$ (in degrees). The emission profile exhibits the structure of a champagne flow of ionized gas (\citealt{tenorio_tagle_1979}). We have marked the location of the ionizing star, $\sigma$ Ori ($r_\m{\star}$), the shock front ($r_\m{s}$), and the rarefaction wave ($r_\m{rw}$). See Sec. \ref{sec:smallershells} for a full description of the champagne flow structure. {\bf (b \& c)} Cuts through Barnards's loop \citep[BL$_\m{top}$ and BL$_\m{bot}$;][]{heiles_2000}, revealing a single peak most prominently seen in H$\alpha$. The dashed lines on the bottom define the bins where the mean optical depth at 353 GHz, $\tau_\m{353,obs}$, from the {\em Planck} dust model R1.20 \citep{abergel_2013} is measured (see text): `BL' is the bin for Barnard's Loop, `BG' is the bin for the background determination. {\bf (d)} Cut through the western edge of the superbubble, coincident with the Eri C filament. Here, the gas and dust reveal a stratified structure, where the ionized gas peaks inside the dust shell.}
\label{fig:cuts}
\end{figure*}

Figure \ref{fig:cuts} shows cross-cuts through the Orion-Eridanus superbubble, revealing the gas- and dust structures of the superbubble and its interior components. Note that Barnard's Loop peaks dramatically in the ionized gas, while in the dust and neutral hydrogen it is only seen as a modest increase in the {\em Planck}, {\em WISE}, and LAB channels (Fig. \ref{fig:cuts}a,b,c). At this point, we find no evidence that a dense shell encapsulates the ionized portion of Barnard's Loop, which presence was predicted by \citet{cowie_1979} (Sec. \ref{sec:discussion}). In contrast, the cross-cut shown in Fig. {\ref{fig:cuts}d} that traces the western bubble shell at high latitudes reveals a stratified structure, where a dust/\HI shell is clearly observed to encapsulate the Eridanus C filament. The absence of an HI shell associated with Barnard's Loop is surprising, and encouraged us to investigate the gas- and dust content of Barnard's Loop and the Orion-Eridanus superbubble in more detail; our findings are presented in the next subsections.

\subsection{Dust in Barnard's Loop}\label{sec:bl}

\subsubsection{Optical depth}\label{sec:blopt}

\citet{heiles_2000} defined two small regions of Barnard's Loop which are clear of any dense molecular clouds along the line of sight; we use the same apertures in the following analysis ('BL$_\m{top}$' and 'BL$_\m{bot}$'; Fig. \ref{fig:superbubble}) to characterize the dust contained in Barnard's Loop. The distance toward Barnard's Loop is uncertain; here, we will assume that Barnard's Loop lies at the distance of the Orion nebula (400 pc; \citealt{sandstrom_2007,menten_2007}). Electron densities inside Barnard's Loop are estimated at $n_\m{e}$ = 2.0 cm$^{-3}$ \citep{heiles_2000}, and at $n_\m{e}$ = 3.2 cm$^{-3}$ from the surface brightness of the H$\beta$ line \citep{o'dell_2011}. A similar analysis of the H$\alpha$ surface brightness here leads to $n_\m{e}$ = 3.4 cm$^{-3}$, after correcting for limb-brightening (a factor 7; \citealt{o'dell_2011}) of a constant density shell with an apparent radial thickness of 1$^\circ$, which amounts to 7 pc at the distance of 400 pc.

Using $n_\m{e}$ = 3.4 cm$^{-3}$ and a radial thickness of 7 pc, adopting a normal gas-to-dust ratio of $N_\m{H}$/$A_\m{V}$ = 1.9 $\times$ 10$^{21}$ cm$^{-2}$ magnitude$^{-1}$ \citep{bohlin_1978}, where $N_\m{H}$ is the column density of hydrogen and $A_\m{V}$ the visual extinction, and a UV-to-visual dust absorption ratio of 1.8 \citep{tielens_1985}, the radial dust optical depth for UV photons with energies between 6 and 13.6 eV is $\tau_\m{UV}$ = 0.07. 

Through a modified-blackbody fit to the {\em Planck} data, the {\em Planck} R1.20 dust model provides all-sky maps of dust optical depth (at 353 GHz), temperature, spectral index, and the total dust emission integrated over frequency. We obtain an observed optical depth of Barnard's Loop at 353 GHz, $\tau_\m{353,obs}$, by correcting the value of the optical depth from the Planck model at the location of Barnard's Loop for a background value. For this method we use the bins defined in Figs. \ref{fig:cuts}b\&c: as Barnard's Loop is not encapsulated by a dense shell (Fig. \ref{fig:cuts} and Sec. \ref{sec:apertures}), we determine the background in a region just outside of the ionized portion of Barnard's Loop (which is significant: $\sim$ 60\% of the total value for both regions). In this way, we isolate the dust grains contained within the ionized part of Barnard's Loop. Then, the {\em observed} optical depths are $\tau_\m{353,obs}$ = 9.0 $\times$ 10$^{-6}$ (for BL$_\m{top}$) and $\tau_\m{353,obs}$ = 5.1 $\times$ 10$^{-6}$ (BL$_\m{bot}$). We can compare this to the {\em expected} optical depth, $\tau_\m{353,cal}$, from the density (3.4 cm$^{-3}$), estimated line of sight depth (45 pc for a 1$^{\circ}$ thickness) and the opacity at 353 GHz ($\sigma_\m{\nu}$ = 7.9 $\times$ 10$^{-27}$ cm$^{2}$ H$^{-1}$) appropriate for the diffuse ISM \citep{abergel_2013}. We calculate $\tau_\m{353,cal}$ = 4.0 $\times$ 10$^{-6}$, in close agreement with $\tau_\m{353,obs}$, which implies that dust mixed with the ionized shell of Barnard's Loop is alike to dust seen in the diffuse ISM.

\subsubsection{Temperature}\label{sec:lya}

\citet{heiles_2000} concluded that grain temperatures in Barnard's Loop are higher than in \HI regions, and proposed that Ly$\alpha$ photons might contribute to the heating of the dust. However, direct heating from the radiation field of Orion OB1 was not considered in their study. We derive an intensity of $G_\m{0}$ = 14 reaching Barnard's Loop in units of the Habing field \citep{habing_1968}, taking into account the radiation from the 7 most massive stars in Orion OB ($\theta$ Ori C, $\delta$ Ori, $\iota$ Ori, $\theta$ Ori A, $\sigma$ Ori, $\epsilon$ Ori, $\zeta$ Ori), using stellar parameters from \citet{martins_2005}. However, \citet{o'dell_2001} shows that radiation from the Orion nebula ($\theta$ Ori C and $\theta$ Ori A) might not reach Barnard's Loop, as the nebula is optically thick to ionizing radiation in all directions, except possibly to the southwest. This would decrease $G_\m{0}$ to 11. 

The observed dust temperature from the {\em Planck} model in BL$_\m{top}$ and  BL$_\m{bot}$ are $T_\m{d}$ = 18.7 K and $T_\m{d}$ = 20.5 K. These values are consistent with temperatures found by \citet{heiles_2000}, $T_\m{d}$ = 19.6 K and $T_\m{d}$ = 20.3 K for BL$_\m{top}$ and BL$_\m{bot}$, respectively, and the mean value $T_\m{d}$ = 19.7 K for the entire sky \citep{abergel_2013}, even though $G_\m{0}$ in BL is significantly raised above the average interstellar radiation field of the solar neighborhood (= 1.7 Habing field; \citealt{tielens_2005}). For comparison, at $G_\m{0}$ = 14 a `typical' 0.1 $\mu$m silicate (graphite) grain would be heated towards $T_\m{d}$ = 19.2 (22.5) K \citep{tielens_2005}. At $G_\m{0}$ = 11, this would be $T_\m{d}$ = 18.6 (21.8) K. In this respect, \citet{abergel_2013} and \citet{aniano_2014} argue that observed dust temperatures do not simply trace the intensity of the radiation field, but in addition reflects changes in grain properties, such as the size distribution, grain structure, and material changes. In any case, for both BL$_\m{top}$ and BL$_\m{bot}$, we do not see the necessity of an extra heating source to account for the observed grain temperatures, such as Ly$\alpha$ heating \citep{heiles_2000}. 

\subsection{Dust and gas in the Orion-Eridanus superbubble}\label{sec:apertures}

\subsubsection{Global observational characteristics}\label{sec:bubble}

The all-sky surveys listed in Sec. \ref{sec:obs} allow us to determine the global observational characteristics of the Orion-Eridanus superbubble, which is of potential use to the study of more distant superbubbles that lack the spatial information of the Orion-Eridanus region. We use apertures that encompass the entire superbubble, the OMCs, and Barnard's Loop, respectively. For the OMCs and Barnard's Loop, these apertures coincide with the outlines depicted in Fig. \ref{fig:superbubble}d. While the full extent of the Orion-Eridanus superbubble will be thoroughly discussed in the remainder of this paper, here we already choose the `entire superbubble aperture' to encompass the majority of the H$\alpha$ emission observed in Fig. \ref{fig:superbubble}b, while tracing the outside of HI filaments seen at -40 km s$^{-1}$ $\textgreater$ v$_\m{LSR}$ $\textgreater$ 40 km s$^{-1}$ (Fig. \ref{fig:superbubble}c), corresponding to the velocity extent in which the Orion-Eridanus superbubble has previously been identified \citep{brown_1995}. We measure the H$\alpha$ luminosity and total far-IR to sub-mm luminosity contained within the apertures, which are compared with stellar parameters of Orion OB1.

For the entire superbubble, we proceed as follows. The north- and east side of the superbubble aperture extend close to the Galactic plane and high-latitude clouds. The encompassed IR emission will therefore be contaminated by emission along the line of sight not related to the superbubble. While disentangling the separate contributions to the IR is not straightforward, we attempt to isolate the emission from our region of interest by measuring the average far-IR integrated intensity, $I_\m{IR,fil}$, within the aperture labeled as `IR fil.' (Fig. \ref{fig:superbubble}d), i.e., the dusty shell that encapsulates the limb-brightened Eridanus B and C filaments known to be associated with the superbubble. The intensities are taken from the {\em Planck} R1.20 dust model \citep{abergel_2013} that uses {\em IRAS} 100 $\mu$m and {\em Planck} 857, 545, and 353 GHz data to determine the total integrated far-IR intensity. Subsequently, we define emission from regions that are much brighter than observed in the limb-brightened filaments, i.e., $I_\m{IR}$ $\geq$ 3$I_\m{IR,fil}$, as {\em not} being part of the superbubble. This procedure provides a mask tracing the regions of high IR brightness towards the Galactic plane, and high-latitude fore- and background clouds well. Note that this procedure filters out the emission from, e.g., the OMCs as well; we ultimately add the contributions of the OMC and Barnard's Loop apertures to get the full IR intensity of the superbubble and the relative contributions of the components (see Tab. \ref{tab:obschar}). For the H$\alpha$ map, the background contamination is much less severe and a masking procedure is not necessary. In this respect, we will argue below that all of the observed H$\alpha$ in Fig. \ref{fig:superbubble}b is caused by photo-ionization from Orion OB1. 

While the masking procedure accounts for the high-IR intensities of contaminating clouds towards the Galactic plane, it does not correct for the diffuse low-level background observed over the entire images. Therefore, we apply a {\em global} background subtraction to both the far-IR and H$\alpha$ maps by measuring the background level within a circular region of 0.4 degree radius at ($l$,$b$) = (229.6, -33.7) {\em outside} of the superbubble aperture that appears to be free of obvious emission features and provides a proper representation of the diffuse background emission averaged over latitude. For both maps, we do not include the $\lambda$ Ori region in our analysis as it appears to be encircled by an ionization-bounded \HII\ region \citep{warren_1977,wall_1996}, and can be regarded as being separated from Orion OB1.

Table \ref{tab:obschar} lists two different values of $L_\m{IR}$. The first value is obtained by filling the masked region with the average IR brightness from the unmasked part of the superbubble aperture. In this way, we get an estimate of the full IR intensity of the entire superbubble if it were unaffected by fore- or background emission along the line of sight. The second value is obtained by setting the masked part to zero, and therefore neglects the masked part of the superbubble. The choice of including or excluding the masked part of the superbubble affects the inferred luminosities from the OMC and Barnard's Loop as well, as these apertures are completely immersed in the masked region of the image, and the chosen intensity of the masked region thus effectively acts as a `global' background for both regions. Below, we will include the masked region in our interpretations, as we believe that the inclusion of this part of the superbubble is essential to the analysis of the total energetics of the region. However, we refer the reader to Table \ref{tab:obschar} for the implications of this choice and its impact on the derived parameters. The total far-IR luminosity is $L_\m{IR}$ = 4$\pi$$I_\m{tot}$$S$, where $I_\m{tot}$ and $S$ are the total integrated intensity and projected surface area of the emitting region. For simplicity, we assume that the entire bubble is located at 400 pc. The measured IR luminosity is compared with stellar parameters of Orion OB1 through $\xi_\m{\star}$ = $L_\m{IR}$/$L_\m{\star}$, where $L_\m{\star}$ is the total luminosity of the Orion OB region (1.7 $\times$ 10$^{6}$ $L_\m{\odot}$, re-evaluated at $d$ = 400 pc, and excluding $\lambda$ Ori; \citealt{warren_1977,wall_1996}). The parameter $\xi_\m{\star}$ defines the fraction of stellar radiation that is captured by the dust and re-emitted in the IR averaged over all solid angles. 

The Ly$\alpha$ photon rate, $N_\m{Ly\alpha}$, is calculated by converting the observed H$\alpha$ photon rate to a Ly$\alpha$ photon rate using the ratio $\alpha_\m{B}$/$\alpha_\m{H\alpha}$, where $\alpha_\m{H\alpha}$ is the effective recombination coefficient of H$\alpha$ (1.31 $\times$ 10$^{-13}$ cm$^{3}$ s$^{-1}$ at $T$ = 6000 K, which is the electron temperature in Barnard's Loop and the Eridanus filaments from \citealt{heiles_2000,o'dell_2011,madsen_2006}, we use this value throughout the superbubble), and $\alpha_\m{B}$ is the total recombination coefficient of hydrogen to all levels but the ground state (2.6 $\times$ 10$^{-13}$ cm$^{-3}$ s$^{-1}$; \citealt{osterbrock_2006}). Case B recombination requires that $N_\m{Ly\alpha}$ equals the total amount of recombinations, which is a quantity that can directly be compared to Orion OB1 through $\xi_\m{ion}$ = $N_\m{Ly\alpha}$/$N_\m{ion}$, where $N_\m{ion}$ is the total number of ionizing photons from Orion OB1 (2.7 $\times$ 10$^{49}$ ph s$^{-1}$, \citealt{o'dell_2011}). This ratio measures the fraction of ionizing photons captured by the gas and converted to Ly$\alpha$. In order to directly compare with the IR luminosity $L_\m{IR}$, we convert $N_\m{Ly\alpha}$ to a luminosity through $L_\m{Ly\alpha}$ = $N_\m{Ly\alpha}$$h\nu_\m{\alpha}$,  where $h\nu_\m{\alpha}$ is the Ly$\alpha$ photon energy, and define $L_\m{ion}$ = $N_\m{ion}$$hv_\m{\alpha}$, such that $\xi_\m{ion}$ = $N_\m{Ly\alpha}$/$N_\m{ion}$ = $L_\m{Ly\alpha}$/$L_\m{ion}$. 

\begin{table}
\centering
\caption{Global observational characteristics of the Orion-Eridanus superbubble}
\begin{tabular}{l|c|c|c|c|c}\hline \hline
Region & $T_\m{d}$ & $L_\m{IR}$ & $\xi_\m{\star}$ & $L_\m{Ly\alpha}$ & $\xi_\m{ion} $ \\ 
 & (K) &  (10$^5$ $L_\m{\odot}$) & &  (10$^5$ $L_\m{\odot}$) & \\ \hline
Superbubble & 19.5 & 7.9 (5.7) $^\m{(a)}$ & 0.47 (0.34) & 1.08 & 0.94 \\
OMC & 17.9 & 2.8 (2.9) & 0.16 (0.17) & 0.16 & 0.14 \\
Barnard's Loop & 19.6 & 0.5 (0.6) & 0.03 (0.04) & 0.20 & 0.17 \\ \hline \hline
\end{tabular}
\tablecomments{Listed are: the dust temperature, $T_\m{d}$; the infrared luminosity, $L_\m{IR}$; fraction of $L_\m{IR}$ to the total luminosity of Orion OB1, $\xi_\star$; luminosity measured from H$\alpha$ and converted to Ly$\alpha$, $L_\m{Ly\alpha}$; fraction of Ly$\alpha$ photons to the total amount of ionizing photons of Orion OB1, $\xi_\m{ion}$. The values between brackets denote the values measured when the masked part of the superbubble is set to zero (see text). $^\m{(a)}$: The total IR luminosity in the superbubble + OMC + Barnard's Loop apertures.}
\label{tab:obschar}
\end{table}

The calculated luminosities, $L_\m{IR}$ and $L_\m{Ly\alpha}$, are denoted in Tab. \ref{tab:obschar}. Table \ref{tab:obschar} shows that half of the total amount of stellar radiation of Orion OB1 is trapped in the superbubble and re-radiated in the IR, $\xi_\m{\star}$ $\sim$ 0.5. The molecular clouds account for about 35\% of the total IR emission of the superbubble; the contribution from Barnard's Loop is negligible. In contrast, $\xi_\m{ion}$ of the OMC (0.14) and Barnard's Loop (0.17) are roughly similar, which is because the Orion nebula and IC 434 emission nebula are currently breaking out of the molecular clouds (Sec. \ref{sec:shells}) and are contained within the OMC aperture. Eventually, all ionizing photons are absorbed within the superbubble aperture ($\xi_\m{ion}$ $\sim$ 1). This implies that, on average, Orion OB1 can provide the necessary ionizing power to illuminate the H$\alpha$ structures detected in Fig \ref{fig:superbubble}c, including the H$\alpha$ filaments that run along the outer edge of the superbubble aperture. Nonetheless, it is possible that individual small-scale structures may still be too bright given their size and distance from Orion OB1 \citep{pon_2014a}. Note that some of these filaments lie outside of what has previously been thought to be the edges of the Orion-Eridanus superbubble \citep{reynolds_1979,heiles_1999, bally_2008, pon_2014b}. The dust temperatures measured inside the superbubble aperture and Barnard's Loop are similar to that observed for the entire diffuse sky by {\em Planck} (see discussion in Sec. \ref{sec:lya}). Inside the OMC, this value is somewhat lower, which is what is expected for dense regions \citep{abergel_2013}.

\subsubsection{Tracing the superbubble structure through dust and gas}\label{sec:specreg}

To investigate the gas- and dust content in specific regions throughout the Orion-Eridanus region, we use apertures depicted in Fig. \ref{fig:superbubble} that are defined such that the majority of H$\alpha$ and far-IR emission from the regions are enclosed (for example, the stratified emission from the limb-brightened emission of the superbubble wall; Fig. \ref{fig:cuts}d), while a low background level is ensured through comparison with the {\em Planck} and LAB maps (the H$\alpha$ emission does not show a high background in general). As the diffuse background emission varies significantly over the entire Orion-Eridanus region, here we use {\em local} background values to properly measure the luminosities contained in the apertures scattered over the Orion-Eridanus region. For the apertures at $b$ $\textgreater$ -30 degrees, we define a background at ($l$,$b$) = (233.2, -28.6) within a circle of 0.4 degree radius. For the apertures at $b$ $\textless$ -30 degrees, we define a background at ($l$,$b$) = (221.9, -49.5). Region 5 is an exception to this rule, as the previously defined background levels were not representative of the local value because of its location amongst high-latitute clouds. For region 5, ($l$,$b$) = (176.3, -15.0) was chosen.

Ly$\alpha$ photons will be resonantly scattered many times in the HII\ region because of its large line cross-section, but eventually be absorbed by dust, contributing to the heating of the dust. The ratio of heating rate by stellar photons, $\Gamma_\m{UV}$, to Ly$\alpha$ photons, $\Gamma_\m{Ly\alpha}$, can be written as \citep{tielens_2005}:

\begin{equation}
\label{eq:heatingone}
\frac{\Gamma_\m{UV}}{\Gamma_\m{Ly\alpha}} = \frac{\pi a^2 n_\m{d} \bar{Q}_\m{abs} L_\m{\star}}{4 \pi r^2 n^2 \alpha_\m{B} h \nu_\m{\alpha}}.
\end{equation} 

\noindent Here, $\pi$$a^2$ and $n_\m{d}$ are the geometrical cross-section and number density of the grains, $\bar{Q}_\m{abs}$ is the average radiation absorption efficiency of the dust, $L_\m{\star}$ is the total luminosity of Orion OB1, $r$ the distance to the source, and $n$ is the hydrogen number density of the gas. Assuming equilibrium between photo-ionization and recombination, for a shell surrounding an empty cavity we have $f_\m{ion}$$N_\m{ion}$ = 4$\pi$$r^2$$n^2$$\alpha_\m{B}$$\Delta$$r$, where $f_\m{ion}$ is the fraction of incident ionizing photons available that are absorbed by the gas locally. Then, Eq. \ref{eq:heatingone} reduces to 

\begin{equation}	
\label{eq:heatingshell}
\frac{\Gamma_\m{UV}}{\Gamma_\m{Ly\alpha}} = \frac{\pi a^2 n_\m{d} \bar{Q}_\m{abs} \Delta r}{f_\m{ion}} \left(\frac{L_\star}{N_\m{ion} h \nu_\m{\alpha}}\right) = \frac{\tau_\m{d}}{f_\m{ion}} \left(\frac{L_\star}{N_\m{ion} h \nu_\m{\alpha}}\right),
\end{equation} 

\noindent where $\tau_\m{d}$ is the (radial) absorption optical depth of the shell. With $L_\star$ = 1.7 $\times$ 10$^{6}$ $L_\m{\odot}$ and $N_\m{ion}$ = 2.7 $\times$ 10$^{49}$ ph s$^{-1}$ (Sec. \ref{sec:bubble}), we have ($L_\star$/$N_\m{ion}$$h\nu_\m{\alpha}$) $\approx$ 15 for Orion OB1, and therefore the ratio $\Gamma_\m{UV}/\Gamma_\m{Ly\alpha}$ equals unity for $\tau_\m{d}$/$f_\m{ion}$ $\approx$ 0.07. 

We can make a priori estimates of $f_\m{ion}$ for the regions defined in Fig. \ref{fig:superbubble}. For Barnard's Loop, we have measured an average $\xi_\m{ion}$ = 0.17 over the entire structure (Table \ref{tab:obschar}). A factor of 0.50 would be expected for a half-sphere geometry if Barnard's Loop were optically thick to ionizing photons. Possibly, a fraction of the ionizing photons from Orion OB1 is trapped within the OMCs before reaching Barnard's Loop, which would lead to $\xi_\m{ion,BL}$ = 0.50 - $\xi_\m{ion,OMC}$ = 0.36 (Table \ref{tab:obschar}). Still, these numbers reveal that 50\% - 66\% of the ionizing photons pass through the half sphere encompassing Barnard's Loop. Hence, we estimate that $f_\m{ion}$ is in the range $\sim$0.33 - 0.50 for the Barnard's Loop apertures BL$_\m{bot}$ and BL$_\m{top}$. The fact that Barnard's Loop is optically thin for the ionizing flux of Orion OB1 can also be inferred from the morphological appearance of the cometary clouds L1617 and L1622, which are located behind Barnard's Loop as measured from the Orion OB association and have sharp I-fronts on their sides facing the Belt stars \citep{bally_2009}. We note that the internal structure of Barnard's Loop may be more complex than that of a homogeneously distributed half-sphere assumed here. A patchy structure containing optically thick clumps ($f_\m{ion}$ $\sim$ 1) and holes through which the photons would leak unhindered ($f_\m{ion}$ $\sim$ 0) could, in principle, lead on average to an $f_\m{ion}$ of order  0.33 - 0.50. However, Barnard's Loop, similar to region 1 and region 6, is not associated with \HI. The remaining regions are associated with neutral hydrogen, where $f_\m{ion}$ will be of order $\sim$ 1 because of the small mean free path of ionizing photons through \HI \citep{tielens_2005}.

\begin{figure*}
\centering
\includegraphics[width=18cm]{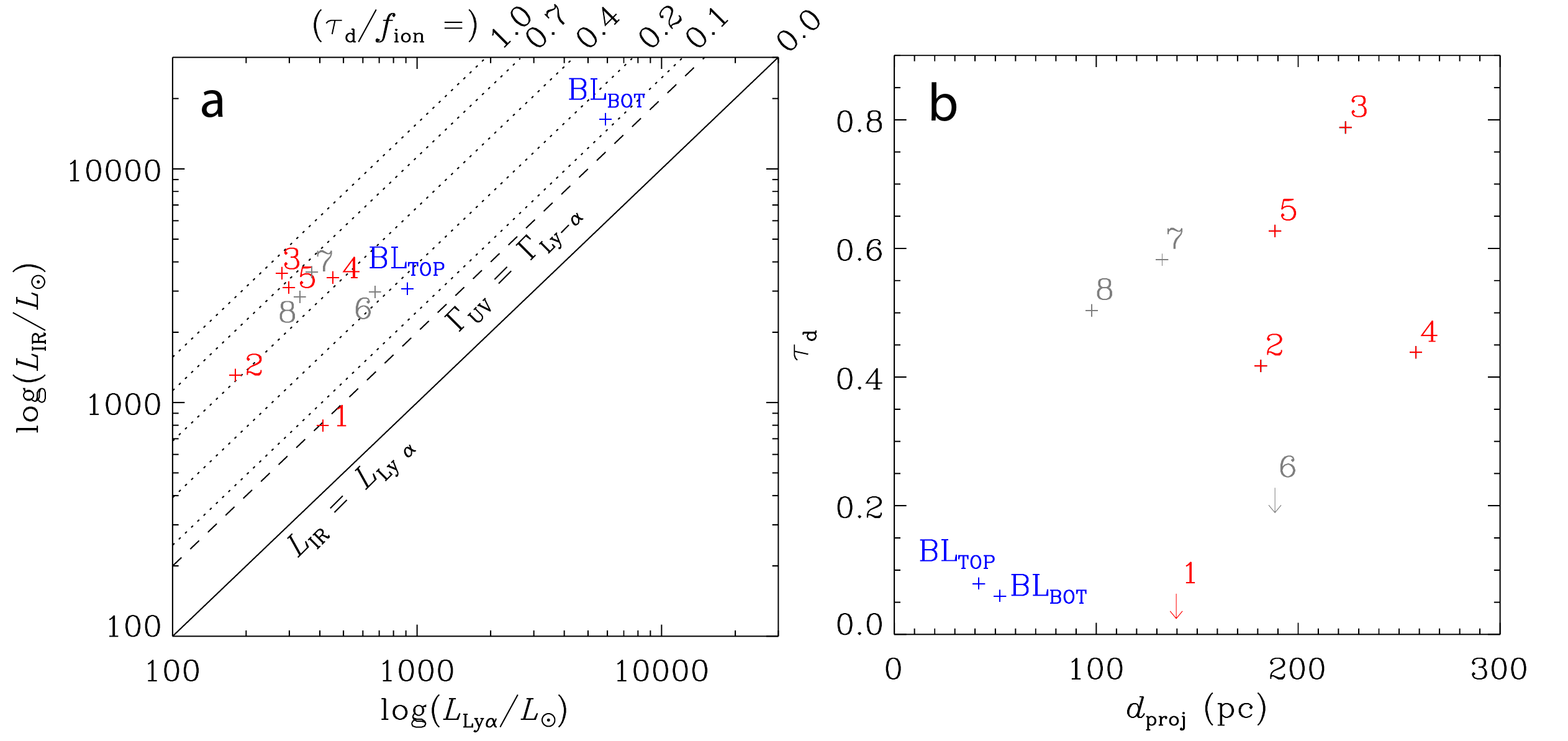} 
\caption{Apertures tracing the structure of the Orion-Eridanus superbubble. {\bf (a)} Infared versus Ly$\alpha$ luminosities measured in the apertures shown in Fig. \ref{fig:superbubble}. The solid line shows where the infrared and Ly$\alpha$ luminosities are equal. The lower dashed line shows where the infrared and Ly$\alpha$ heating rates are equal, corresponding to $\tau_\m{d}$/$f_\m{ion}$ = 0.07 (see text), where $\tau_\m{d}$ us the dust optical depth at UV wavelengths and $f_\m{ion}$ is the fraction of ionizing photons available absorbed by the gas. The dotted lines are similar, but for increasing $\tau_\m{d}$. Overplotted are the luminosities measured in the apertures drawn in Fig. \ref{fig:superbubble}, color-coded as follows: regions 1-5 ({\em red}) are positioned along the projected outer shell of the superbubble aperture (Fig. \ref{fig:superbubble}c); Barnard's loop (BL$_\m{top}$ and BL$_\m{bot}$) ({\em blue}); region 6 - and 8 ({\em gray}) are the Eridanus A filament and molecular clouds close to the Orion OB association. {\bf (b)} The radial dust optical depth at UV wavelengths versus the projected distance towards the H$\alpha$ flux-weighted center of the superbubble.} 
\label{fig:apertures}
\end{figure*}

We measure both $L_\m{IR}$ and $L_\m{Ly\alpha}$ within the apertures shown in Fig. \ref{fig:superbubble}, and compare these measurements with luminosities from Orion OB1. The structures contained in each defined aperture will subtend a solid angle $\Omega$ as measured from Orion OB1. Now, any ionizing photon that is absorbed by hydrogen and converted to Ly$\alpha$ will eventually contribute to the heating of the dust and, therefore, the total IR luminosity that emanates from each region can be written as

\begin{equation}	
\label{eq:iremission}
L_\m{IR} = L_\m{UV} + L_\m{Ly\alpha} =  \tau_\m{d}L_\m{\star} \Omega_\m{UV} + f_\m{ion}N_\m{ion} h \nu_\m{\alpha} \Omega_\m{Ly\alpha},
\end{equation}

where $L_\m{UV}$ is the dust luminosity provided by direct absorption of stellar photons. In the classical picture of an expanding superbubble, a dense, neutral shell will wrap around the 10$^4$ K region created by the reverse shock traced by bright H$\alpha$ emission \citep{weaver_1977}. Therefore, we can assume that the IR and H$\alpha$ trace structures which measure the same solid angle with respect to Orion OB1, $\Omega_\m{UV}$ = $\Omega_\m{Ly\alpha}$, and we can write

\begin{equation}	
\label{eq:taudust}
\frac{(L_\m{IR}/L_\m{\star})}{(L_\m{Ly\alpha}/L_\m{ion})} =  \frac{\tau_\m{d}}{f_\m{ion}} + \left(\frac{N_\m{ion} h \nu_\m{\alpha}}{L_\m{\star}}\right),
\end{equation}

which describes the ratio between the amount of energy that is captured by the dust and re-emitted in the IR, to the amount of energy that is captured by the gas and converted to Ly$\alpha$.

Our results are shown in Fig. \ref{fig:apertures}. All studied regions show $L_\m{IR}$ $\textgreater$ $L_\m{Ly\alpha}$ (Fig. \ref{fig:apertures}a), which reveals that a (substantial) amount of stellar radiation is directly absorbed by the dust. Here we note that $L_\m{IR}$ = 2$L_\m{Ly\alpha}$ when $\Gamma_\m{UV}$ = $\Gamma_\m{Ly\alpha}$ (cf. Eq. \ref{eq:iremission}). An interesting trend can be seen from Fig. \ref{fig:apertures}a. The regions 2 - 5, positioned on filamentary structures seen in gas and dust along the outside of the superbubble aperture depicted in Fig. \ref{fig:superbubble}, reveal significant higher fractions of $\tau_\m{d}$/$f_\m{ion}$ compared to the regions located in Barnard's Loop (BL$_\m{bot}$, BL$_\m{top}$). This implies that stellar photons, provided by Orion OB1 (Sec. \ref{sec:bubble}), are more efficiently absorbed in regions 2 - 5 compared to that in Barnard's Loop. In other words, the dust optical depth for stellar photons associated with Barnard's Loop is very small. In Fig. \ref{fig:apertures}b, we rewrite Eq. \ref{eq:taudust} for $\tau_\m{d}$/$f_\m{ion}$ and, together with our estimations for $f_\m{ion}$ as described above, plot the derived dust optical depth in the regions as a function of projected distance $d_\m{proj}$ from ($l$,$b$) = (206.5,-18.0), the flux-weighted center of the bubble as measured from the H$\alpha$ emission (see Fig. \ref{fig:superbubble}b; \citealt{reynolds_1979}). We exclude $\lambda$ Ori as it seems to be encircled by an ionization-bounded \HII region. Still, in Sec. \ref{sec:lambdaori} we will show that the ionized gas in the $\lambda$ Ori region shows the characteristics of a champagne flow \citep{tenorio_tagle_1979}, in which case the \HII region will be density bounded on the side of the outward champagne flow, and radiation will likely be able to escape the region. However, the circular morphology of the $\lambda$ Ori bubble detected in H$\alpha$ (Fig. \ref{fig:wham}) reveals that the champagne phase, if present, may not have fully developed (Sec. \ref{sec:lambdaori}) and, therefore, we will assume that the $\lambda$ Ori \HII region is still ionization-bounded.

For the BL apertures, we plot the results for $\tau_\m{d}$ versus $d_\m{proj}$  assuming $f_\m{ion}$ = 0.5 (Sec. \ref{sec:specreg}). For regions 1 and 6, the absence of a clear neutral HI filament prohibited us from predicting $f_\m{ion}$, and only an upper limit to $\tau_\m{d}$ is given. The results plotted in Fig. \ref{fig:apertures}b, albeit suffering from low-number statistics, indicate a trend where $\tau_\m{d}$ rises with increasing $d_\m{proj}$.

From Fig. \ref{fig:apertures}, it becomes apparent that dust associated with Barnard's loop is not able to capture all of the radiation emitted by the Orion OB1 association. In this context, the bright H$\alpha$ filament that surrounds Barnard's Loop, from region 1 up to the Galactic plane (Fig. \ref{fig:superbubble}), is lit up by ionizing photons breaking through Barnard's Loop. A straight line from the flux-weighted centre of the superbubble (Fig. \ref{fig:superbubble}), through Barnard's Loop to the outer filament in region 1, reveals a flux ratio of $F_\m{BL}$/$F_\m{reg. 1}$ $\sim$ 5, where $F_\m{BL}$ and $F_\m{reg. 1}$ are the measured peak H$\alpha$ fluxes in Barnard's Loop and the outer filament in region 1, respectively. At a distance ratio of $d_\m{BL}$/$d_\m{reg. 1}$ $\sim$ 2.5, this is consistent with both the structures absorbing a similar amount of ionizing photons, confirming $f_\m{ion}$ $\sim$ 0.5 for the Barnard's Loop apertures as was assumed in our derivations of $\tau_\m{d}$ above.
 
As a sanity check to our methodology, we also plot in Fig. \ref{fig:apertures} the results for region 6 (corresponding to the southern end of the Eri A filament), which reveals low $\tau_\m{d}$ compared to the other Eridanus filaments. However, the association of the Eridanus A filament with the superbubble is still controversial and is discussed thoroughly in \citet{pon_2014a}. For completeness, we show the results for apertures located on the G203-37 and GS204-31 clouds (region 7 \& 8) which are thought to be {\em inside} and at the near-side of the Orion-Eridanus superbubble \citep{snowden_1995}. The clouds are highly opaque to 0.25 keV radiation with intervening hydrogen column densities of $\sim$8 $\times$ 10$^{20}$ cm$^2$ \citep{snowden_1995}, translating to an optical depth at UV wavelength of $\tau_\m{UV}$ $\sim$ 0.8 using the relations given in Sec. \ref{sec:blopt}, in rough agreement with our measurements shown in Fig. \ref{fig:apertures}. 

With $f_\m{ion}$ = 0.5 for the Barnard's Loop apertures, the dust optical depth as measured from the IR ($\tau_\m{d}$ = 0.06 and $\tau_\m{d}$ = 0.08 for BL$_\m{top}$ and BL$_\m{bot}$, respectively; Fig. \ref{fig:apertures}a \& b) is in agreement with a homogeneous shell containing dust similar to that seen in the diffuse ISM ($\tau_\m{UV}$ = 0.07; Sec. \ref{sec:bl}). Finally, we note that while Ly$\alpha$ contributes about 25-50\% of the total energy absorbed by dust in Barnard's Loop (Fig. \ref{fig:apertures}a), this will have little effect on the dust temperature (Sec. \ref{sec:lya}) as $T_\m{d}$ only weakly depends on the heating rate. In particular, $T_\m{d}$ $\propto$ ($\Gamma_\m{UV} + \Gamma_\m{Ly\alpha}$)$^{1/6}$ for silicate grains \citep{tielens_2005}.

\subsection{Expanding shells within the Orion-Eridanus region}\label{sec:shells}

\subsubsection{High- and intermediate velocity gas}\label{sec:hvivgas}

The presence of high-velocity (HV) absorption features of ionized gas between -120 km s$^{-1}$ $\lesssim$ $v_\m{LSR}$ $\lesssim$ -80 km s$^{-1}$ was detected in multiple lines of sight towards Orion \citep{cowie_1979,welty_2002}. In addition, intermediate-velocity (IV) features were detected between -80 km s$^{-1}$ $\lesssim$ $v_\m{LSR}$ $\lesssim$ -20 km s$^{-1}$ \citep{cowie_1978,cowie_1979,huang_1995,welty_2002}. Both components show similar depletions, ion ratios, and physical conditions, albeit the IV gas contains column densities that exceed that of the HV gas by a factor of 3 \citep[e.g.,][]{welty_2002}. The HV gas, dubbed by \citet{cowie_1979} as ``Orion's Cloak", has an angular diameter of at least 15$^\circ$ and is ascribed to a recent supernova that, from its dynamical expansion scale, should have occurred some 3 $\times$ 10$^5$ yr ago. 

\begin{figure*}
\centering
\includegraphics[width=18cm]{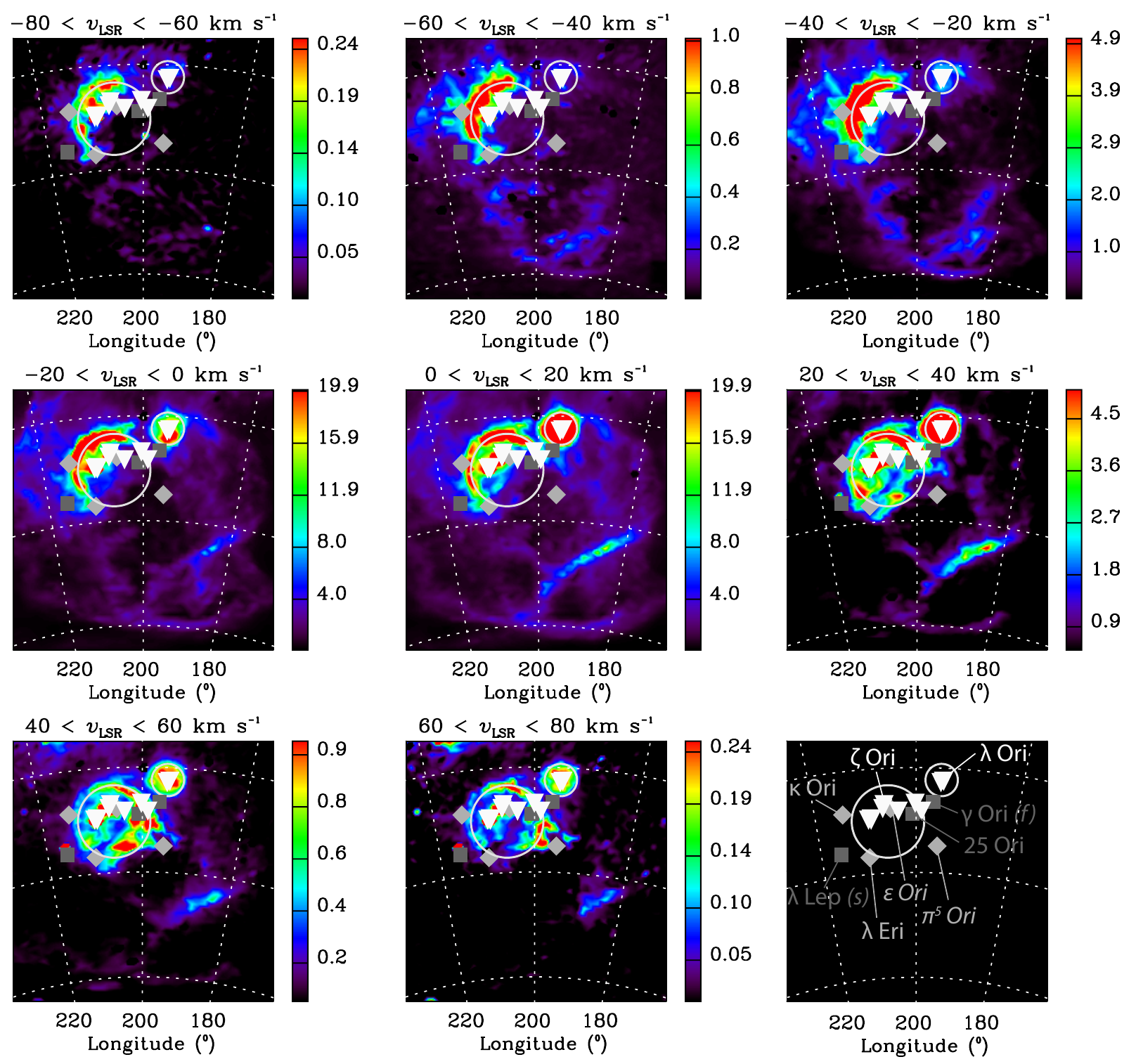} 
\caption{WHAM H$\alpha$ observations of the Orion-Eridanus superbubble. Shown are color-coded intensity maps (in units of Rayleigh) integrated over a velocity range in $v_\m{LSR}$ that is indicated on top of each panel. Note the different scaling of each individual panel. Overplotted are lines of sight from absorption line studies \citep{cowie_1978,cowie_1979,huang_1995,welty_2002} that show {\em both} high-velocity and intermediate-velocity gas {\em (inverted triangles)}, lines of sight that show {\em only} intermediate-velocity gas {\em (diamonds)}, and lines of sight that do not show a clear detection of high-velocity or intermediate-velocity gas  {\em (squares)}. The classes are plotted in different gray scalings to make them discernable, and we have explicitely labeled some of the stars in the lower right panel for orientation. The stars $\pi^5$ Ori and $\epsilon$ Ori are denoted in italics as observations only covered velocities of $|v_\m{LSR}|$ $\textless$ 60 km s$^{-1}$ and $|v_\m{LSR}|$ $\textless$ 100 km s$^{-1}$, respectively \citep{cowie_1979}. Measurements of HV/IV gas in $\lambda$ Lep are hindered \citep{cowie_1979} because of stellar rotation {\em (s)}, while $\gamma$ Ori lies at 77 pc \citep{van_leeuwen_2007} and may trace foreground gas {\em (f)}.}
\label{fig:wham}
\end{figure*}

Figure \ref{fig:wham} shows H$\alpha$ velocity maps from the WHAM H$\alpha$ spectral survey \citep{haffner_2003} of the Orion-Eridanus superbubble, covering the velocity space between -80 km s$^{-1}$ $\lesssim$ $v_\m{LSR}$ $\lesssim$ +80 km s$^{-1}$. Overplotted are sightlines from absorption line studies \citep{cowie_1978, cowie_1979, huang_1995, welty_2002}. The HV and IV gas trace structures around Barnard's Loop and the $\lambda$ Ori ring that extend outside of the traditional limits of the Orion-Eridanus superbubble  \citep{reynolds_1979,heiles_1999,bally_2008,pon_2014b}, as demonstrated by the sightlines of $\kappa$ Ori, $\lambda$ Ori, and $\lambda$ Eri. While it must be recognized that the distribution of the gases are somewhat patchy \citep{cowie_1979}, it is clear that the IV and HV gas trace two distinct components. Moreover, the HV gas seems to be confined to a more limited area compared to the IV gas, as reflected by the sightlines that show {\em only} the IV gas.

We note upfront that electron temperatures within Barnard's Loop and $\lambda$ Ori are determined at $\sim$6000 - 7500 K \citep{reich_1978,heiles_2000,o'dell_2011,madsen_2006} which excludes that the HV gas originates from high-velocity wings of a thermal Gaussian line profile (the thermal broadening can be estimated with $\sqrt{2kT/m_\m{H}}$ $\sim$ 10 km s$^{-1}$, where $k$ is Boltzmann's constant and $m_\m{H}$ is the mass of a hydrogen atom). The velocity information of the Eridanus filaments are discussed in \citet{pon_2014a}. Of particular interest here are the encircled regions towards the east of the superbubble, covering the Barnard's Loop region and the $\lambda$ Ori region, as well as smaller-scale bubbles. 

\subsubsection{Barnard's Loop bubble}\label{sec:blexp}

The intensity peak of Barnard's Loop lies between -20 km s$^{-1}$ $\lesssim$ $v_\m{LSR}$ $\lesssim$ +20 km s$^{-1}$, but the structure is visible throughout the entire velocity space covered by the WHAM survey. Here, we reveal the presence of an ionized filament, visible {\em only} at positive velocities $v_\m{LSR}$ $\textgreater$ +20 km s$^{-1}$. This filament appears to form a complete bubble structure together with the bright crescent of Barnard's Loop, centered on ($l$,$b$) = (193,-20) with radius 14$^{\circ}$. The lines of sight covered by the absorption survey of \citet{cowie_1979} indicate that the HV gas originates from the encircled H$\alpha$ bubble structure, or the `Barnard's Loop Bubble' (Fig. \ref{fig:wham}), suggesting a connection between both components. \citet{cowie_1979} estimated the size of Orion's Cloak at a minimum of 15 degrees that matches the size of the Barnard's Loop bubble. Thus, the Barnard's Loop bubble may well be part of the supernova remnant associated with Orion's Cloak.  
 
The most pronounced radial velocity effects are expected to originate from the center of an expanding bubble. For the HV gas, these velocities lie outside of the velocity range covered by the WHAM survey (-80 km s$^{-1}$ $\lesssim$ $v_\m{LSR}$ $\lesssim$ 80 km s$^{-1}$;  \citealt{haffner_2003}), rendering WHAM unable to pick up the HV gas detected by {\em Copernicus} \citep{cowie_1979} and {\em HST} \citep{welty_2002}, although it is questionable if WHAM would be able to pick up the low surface brightness of the HV gas expected in H$\alpha$ ($\sim$3 $\times$ 10$^{-3}$ Rayleigh; \citealt{cowie_1979}). The OMCs are roughly centered on $v_\m{LSR}$ = 10 km s$^{-1}$ in velocity-space \citep{wilson_2005}; if we assume a systemic velocity towards the Orion region of $v_\m{LSR}$ = 10 km s$^{-1}$ and connect the HV gas with the expansion of the Barnard's Loop bubble, we estimate its expansion velocity at $v_\m{exp}$ = 100 km s$^{-1}$. However, it is not clear how much of the mass within the Barnard's Loop bubble (Sec. \ref{sec:energetics}) is associated with this expansion velocity. The coverage of sightlines of the Orion-Eridanus region is limited, and the HV gas could in principle trace small, high-velocity components related to the Barnard's Loop bubble, rather than tracing the bulk expansion of the bubble itself. Here we implicitly assume that the bulk of the gas in the expanding shell is expanding at the adopted expansion velocity. The pre-shock densities for the HV gas towards $\zeta$ Ori ($n_\m{0}$ = 3 $\times$ 10$^{-3}$ cm$^{-3}$; \citealt{welty_2002}) then implies that the bubble expands either into the Hot Intercloud Medium (HIM; $T$ $\sim$ 10$^6$ K, $n$ $\sim$ 3 $\times$ 10$^{-3}$ cm$^{-3}$) that pervades the Galactic Halo, or a pre-existing cavity, such as the superbubble interior. We defer a thorough discussion of the Barnard's Loop bubble and its connection to the HV gas to section \ref{sec:discussion}.

\subsubsection{$\lambda$ Ori bubble}\label{sec:lambdaori}

\begin{figure}
\centering
\includegraphics[width=8.75cm]{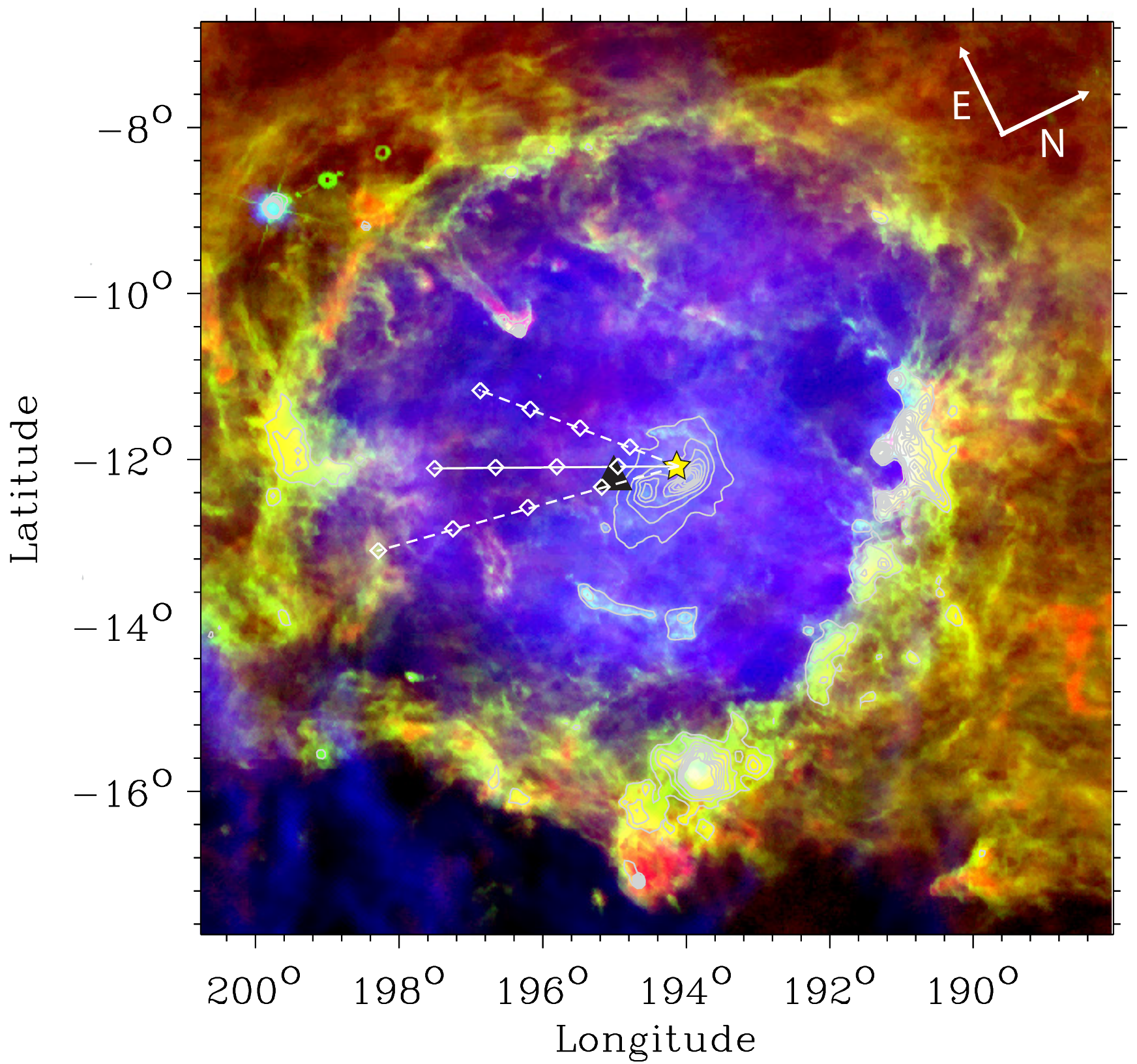} 
\caption{Three-color image of the $\lambda$ Ori bubble. The color codes are the same as Fig. \ref{fig:superbubble}a. Overlayed are contours of IRAS 60 $\mu$m, with intensity levels: maximum - 90\% - 80\% - 70\% - 60\% - 50\% - 40\% - 30\%. The solid line marks the past trajectory of the star $\lambda$ Ori during 4 Myr given its current space motion. Open diamond symbols mark intervals of 1 Myr. The dashed lines are the extremes of possible space motion considering the errors in proper motion and distance to the star. The black triangle marks the centre of expansion of the $\lambda$ Ori bubble \citep{lang_1998}.}
\label{fig:lori}
\end{figure}

The circular-symmetric \HII\ region surrounding $\lambda$ Ori has been known for a long time. Through a study of the star formation history of the $\lambda$
 Ori region, \citet{dolan_1999} hypothesized that about 1 Myr ago, a supernova disrupted the parent molecular cloud and created the giant \HII region seen to date, which is maintained by the remaining population of massive stars in the region. The expansion velocity of the \HII region is traced by observations of the dense molecular shell surrounding the ionized gas, and is estimated at  $v_\m{exp}$ =16.5 km s$^{-1}$ expanding from the point ($l$,$b$) = (195.8,-12.1) centered at $v_\m{LSR}$ = 3.8 km s$^{-1}$ in velocity space \citep{lang_1998}.
 
The main ionizing star $\lambda$ Ori lies right at the (projected) heart of the bubble, which seems inconsistent with its proper motion ($\mu_{\alpha}$\,cos\,$\delta$ = -0.34$\pm$0.60 mas and $\mu_{\delta}$ = -2.94$\pm$0.33 mas, \citealt{van_leeuwen_2007}) with respect to the dense, molecular shell \citep{maddalena_1987,lang_1998}. We calculate the LSR space velocity of $\lambda$ Ori following the method described in \citet{cox_2012}, using its proper motion  \citep{van_leeuwen_2007}, radial velocity ($v_\m{rad}$ = 30.1 km s$^{-1}$; \citealt{gontcharov_2006}), while correcting for the solar motion using parameters from \citet{coskonoglu_2011}. We adopt a distance of 450 pc to the star \citep{dolan_2001}. The calculated space motion of $\lambda$ Ori is $v_\m{rad}$ = 18.2 km s$^{-1}$ with a position angle of 330 degrees (from north-to-east) inclined by 55 degrees in the plane of the sky. The trajectory is overplotted in Fig. \ref{fig:lori}, and reveals that it is unlikely that its current motion has been imparted at birth, since $\lambda$ Ori would have then originated at the boundary between the molecular and ionized gas (the expected age of $\lambda$ Ori is $\sim$3 - 5 Myr; see \citealt{hernandez_2010}, and references therein). However, its motion and distance from the center of expansion (Fig. \ref{fig:lori}) {\em is} consistent with the SN hypothesis that would have launched $\lambda$ Ori on its current trajectory about 1 Myr ago. 
 
We recognize the arc-shaped mid-infrared emission around $\lambda$ Ori, which is the tell-tale signature of a dust wave, created by the interaction of dust entrained in a champagne flow with radiation pressure from the star, which is a common phenomena inside interstellar bubbles \citep{ochsendorf_2014a,ochsendorf_2014b}. Indeed, the radio continuum maps presented by \citet{reich_1978} reveal an emission gradient within the ionized gas that is expected for a champagne flow \citep{tenorio_tagle_1979}. The emission gradient and the location of the dust wave reveals that the interior of the $\lambda$ Ori bubble is venting into the surrounding medium towards the south-east. However, the overall distribution of the ionized gas is still roughly spherical (Fig. \ref{fig:wham}); an extended emission region that would accompany a developed \HII region champagne flow is not detected. This may indicate that the champagne phase has just commenced and the bubble is on the verge of breaking out. The implications of these observations shall be addressed in Sec. \ref{sec:rejuvenation}.

The WHAM observations in Fig. \ref{fig:wham} reveal that the $\lambda$ Ori \HII\ region lights up at positive $v_\m{LSR}$ (Fig. \ref{fig:wham}), extending longwards of $v_\m{LSR}$ $\textgreater$ 80 km s$^{-1}$. In addition, elevated emission surrounding the $\lambda$ Ori region can be seen at the other velocity extreme of WHAM, i.e., -80 km s$^{-1}$ $\lesssim$ $v_\m{LSR}$ $\lesssim$ -60 km s$^{-1}$, possibly tracing the same component from \citet{cowie_1979}, who detected faint HV gas features towards $\lambda$ Ori. The FWHM of the H$\alpha$ line in the $\lambda$ Ori region is $\sim$80 km s$^{-1}$, much broader than the thermal linewidth ($\sim$ 10 km s$^{-1}$; Sec. \ref{sec:hvivgas}). Thus, both the HV gas and H$\alpha$ suggests higher velocities than that observed expansion velocity of the molecular shell ($v_\m{exp}$ = 16.5 km s$^{-1}$; \citealt{lang_1998}). The discrepancy between velocities observed for the dense molecular ring and the ionized gas could indicate a fast blister flow (see above) that can lead to large line of sight variations, or a recent explosive event that has accelerated the gas in the bubble interior, but has not yet coupled to the dense molecular ring.

\subsubsection{GS206-17+13 and the Orion nebula}\label{sec:smallershells}
Besides the obvious large-scale shells that are associated with Barnard's Loop and $\lambda$ Ori, we note that the Orion complex harbors shell structures on all detectable scales. Below, we will discuss the degree-sized GS206-17+13 shell \citep{ehlerova_2005} and the shell surrounding the Orion Nebula \citep{o'dell_2001,guedel_2008}, but on even smaller scales there have been detections of many regions within the Orion A cloud that appear to have shell-like structures, presumably driven by \HII regions from massive young stellar objects \citep{bally_1987,heyer_1992}. Both GS206-17+13 and the Orion Nebula emit at positive $v_\m{LSR}$ (Fig. \ref{fig:wham}) with large line-widths (FWHM $\sim$70 km s$^{-1}$) that can both be related to ionized (blister) flows as described below.

\citet{ochsendorf_2015} discussed the GS206-17+13 shell which is directly adjacent to the OMC B, has dimensions of 2$^{\circ}$ $\times$ 4$^{\circ}$ (14 $\times$ 28 pc), and is approximately centered on the $\sigma$ Ori star cluster. The formation of the shell is likely caused by the stellar winds of the Orion OB1b subgroup that formed the protrusion out of the OMC B. The bright \HII emission that is projected inside part of the shell (the IC 434 emission nebula) is not related to the formation or history of GS206-17+13, as this gas originates from the champagne flow driven by the ionizing flux of the $\sigma$ Ori cluster that has entered the pre-existing bubble cavity and is now approaching OMC B \citep{ochsendorf_2015}. Ionization of the cloud edge sets up a pressure discontinuity between the cloud and inter cloud medium, and drives a shock front into the low-density bubble, while a rarefaction wave travels into the OMC B cloud. In panel 2a of Fig. \ref{fig:cuts}, the shock front is visible at $r_\m{s}$, whereas the rarefaction wave is located at $r_\m{rw}$. Between the shock fronts, a champagne flow of ionized gas is set up \citep{tenorio_tagle_1979}, flowing from $r_\m{rw}$ towards $r_\m{s}$. 

The density structure of the champagne flow of the IC 434 region is well-characterized by an exponential density gradient \citep{ochsendorf_2015}, reaching from $n_\m{H}$  = 35 cm$^{-3}$ at the IF on the OMC B cloud surface, to 3.5 cm$^{-1}$ at the location of $\sigma$ Ori ($r_\m{\star}$; Fig. \ref{fig:cuts}) where the flow reaches $\sim$30 km s$^{-1}$ (Mach number $\mathcal{M}$ $\simeq$ 3 for a 10$^4$ K gas), which is the typical maximum velocity reached for champagne flows \citep{tenorio_tagle_1979}. Therefore, the ionized gas will not accelerate much further (also reflected by the flattening of the emission profile behind $r_\m{\star}$; Fig. \ref{fig:cuts}) and the density in the flow remains roughly similar up until the shock front, i.e., $\rho_\m{1}$ $\simeq$ 3.5 cm$^{-3}$. Then the pre-shock density $\rho_\m{0}$ in which the shock front is moving can be determined through $\rho_\m{1}$/$\mathcal{M}^2$ = $\rho_\m{0}$ \citep{bedijn_1981}, from which we infer $\rho_\m{0}$ = 0.4 cm$^{-3}$, which is far below that observed for electron densities within evolved Galactic \HII regions of similar size ($\gtrsim$ 50 cm$^{-3}$; \citealt{paladini_2012}), however typical for densities observed within the WIM \citep{tielens_2005}. Moreover, the existence of the cometary clouds L1617 and L1622, which are located behind Barnard's Loop as measured from the Orion OB association and have sharp I-fronts on their sides facing the Belt stars \citep{bally_2009} reveal that the ionizing flux of the Belt stars must escape the GS206-17+13 shell, from which we conclude that the shell is incomplete and the bubble has likely `burst'. Hence, we posit that the bubble interior has been emptied in its past, and its contents was channeled into the Orion-Eridanus superbubble cavity.

The Orion nebula is a prototypical blister \HII region created by the young stars of the Trapezium  ($\textless$1 Myr; \citealt{hillenbrand_1997,bally_2008}) with $\theta^1$ Ori C as the dominant ionizing source. The ionized gas is streaming from the near-side of the OMC at about 10 km s$^{-1}$ \citep{o'dell_2001} and accelerates into the surrounding medium. Recent velocity measurement of $\theta^1$ Ori C reveal that the star may be moving away from the OMC at a velocity of 13 km s$^{-1}$ (see \citealt{o'dell_2009}, and references therein), which would imply a dynamical timescale of the Orion nebula of only 1.5 $\times$ 10$^4$ yr, since the star is currently located $\sim$0.2 pc away from the IF. Indeed, the measured EM reveals that the thickness of the ionized gas layer is only 0.13 pc \citep{wen_1995} assuming a constant density, which is known to be $\sim$6 $\times$ 10$^3$ cm$^{-3}$ at the IF \citep{o'dell_2001}. X-ray emission from the $\sim$10$^6$ K gas of the shocked stellar wind of $\theta$ Ori C \citep{guedel_2008} has been detected in the Extended Orion Nebula (EON) and confirms the  leakage of the (hot) gas towards the south-west, with the Orion-Eridanus superbubble as the most likely outlet of the flow of material.

\section{Discussion}\label{sec:discussion}
		
\citet{cowie_1979} interpreted neutral absorption lines at $|v_\m{LSR}|$ $\lesssim$ 20 km s$^{-1}$ towards stars in Orion as a thick, dense shell of swept-up material, partially ionized on the inside by the Orion OB association. This ionized component would be evident as the IV gas and Barnard's Loop. However, the LV and IV gas are clearly separated in velocity space \citep{welty_2002} towards $\zeta$ Ori at ($l,b$) = (206.5,-16.6). This velocity separation makes it unlikely that LV and IV trace the same component of the Orion-Eridanus superbubble. In this respect, \citet{bally_2008} already noted that the velocity dispersion of the ensemble of atomic and molecular clouds in Orion ranges over $\textgreater$ 20 km s$^{-1}$. Therefore, we attribute the low-velocity neutral gas to clouds in the Orion region instead of a dense, neutral shell surrounding Orion. In addition, we have revealed the absence of a dense neutral shell surrounding Barnard's Loop (Sec. \ref{sec:apertures}), and have shown that the Loop is part of a (separate) complete shell structure that is spatially correlated with HV gas detected towards Orion \citep{cowie_1979,welty_2002} as discussed in Sec. \ref{sec:shells}.

The above described results call for a thorough revisit of the structure, history, and subsequent evolution of the Orion-Eridanus superbubble. Below in Sec. \ref{sec:absorption}, we will connect the different gas tracers to the components of the superbubble, and combine this with our observational work in order to derive a new and improved picture of the Orion-Eridanus region in Sec. \ref{sec:extent} \& Sec. \ref{sec:tempint}. We relate our findings to the stellar content of Orion OB1 in Sec. \ref{sec:relationob1} and present a scenario for the future evolution of the Orion-Eridanus superbubble in Sec. \ref{sec:rejuvenation}.

\subsection{Connecting the dots: an updated picture of the Orion-Eridanus superbubble}\label{sec:connectingdots}

\subsubsection{Probing the superbubble through absorption lines}\label{sec:absorption}

Intermediate-velocity gas components are detected over a wide wavelength range, i.e., -80 km s$^{-1}$ $\lesssim$ $v_\m{LSR}$ $\lesssim$ -20 km s$^{-1}$ \citep{cowie_1978,cowie_1979,huang_1995,welty_2002}. However, with high spectral resolution, high signal-to-noise, and broad spectral coverage data, \citet{welty_2002} were able to locate the bulk of the IV gas towards $\zeta$ Ori in two single components at $v_\m{lsr}$ = -35.3 km s$^{-1}$ and $v_\m{lsr}$ = -31.6 km s$^{-1}$.

We link the IV gas to the outer superbubble wall through several observations. First, the bulk of the IV gas lies within the velocity space reported for the superbubble wall in \HI (-40 km s$^{-1}$ $\lesssim$ $v_\m{LSR}$ $\lesssim$ 40 km s$^{-1}$; \citealt{brown_1995}). We note that by connecting the (ionized) IV to the superbubble wall, we assume that the outer wall towards $\zeta$ Ori is predominantly ionized along the line of sight. This might connect with our result that the eastern part of the superbubble wall does not exhibit a neutral shell, as was observed in region 1 of Fig. \ref{fig:superbubble} and will be further discussed in Sec. \ref{sec:tempint}. Second, the total \HII\ column density (log[N(\HII)] = 18.44 cm$^{-2}$) and the post-shock conditions of the IV gas towards $\zeta$ Ori ($n_\m{1}$ = 0.16 cm$^{-3}$, $T_\m{1}$ = 9e3 K; \citealt{welty_2002}) imply an ionized shell thickness of $\sim$5.6 pc, in close agreement with the measured thickness of 1 degree or $\sim$7 pc of the H$\alpha$ filaments along regions 1 - 5 in Fig. \ref{fig:superbubble}. Third, the IV gas is widespread, extending well outside Barnard's Loop and encompassing the $\lambda$ Ori region, tracing a distinct component over an area larger than the HV gas (Sec. \ref{sec:hvivgas}). Even though the definite extent of the IV gas should be established through an absorption line study that include sightlines covering the entire region shown in Fig. \ref{fig:wham}, here we argue that that the IV gas traces a structure significantly larger than the dimensions of the Orion-Eridanus superbubble described in previous works \citep{reynolds_1979,heiles_1999,bally_2008,pon_2014b}.

The above results reveal that the IV gas may be linked to the outer superbubble wall, whose neutral component was defined in \citet{brown_1995}. In Sec. \ref{sec:shells}, we have argued that the HV gas traces the fast expansion of the Barnard's Loop bubble through the correlation between the sightlines that exhibit HV features and the morphology of the Barnard's Loop bubble from H$\alpha$ (Fig. \ref{fig:wham}). A similar result was obtained for the $\lambda$ Ori bubble. Both the Barnard's Loop bubble and the $\lambda$ Ori bubble may be connected to recent SN explosions (Secs. \ref{sec:blexp} \& \ref{sec:lambdaori}) that could be at the origin of the HV gas. Finally, the LV gas is attributed to the ensemble of clouds in Orion \citep{bally_2008}. As a caveat, we note that these conclusions are based on the study of a limited amount of sightlines that may not fully reflect the complexity of the Orion region (see Sec. \ref{sec:summaryextent}). Nonetheless, based on the currently available data, we hypothesize that the different gas tracers trace separate components of the superbubble, summarized in Table \ref{tab:connection}. 

\begin{table*}
\centering
\caption{The connection between observed gas velocities and components within the Orion-Eridanus region}
\begin{tabular}{l|c|c|c|c|c|c|l}\hline \hline
Component & $v_\m{LSR}$ & $n_\m{0}$ &  $n_\m{1}$ & log[$N_\m{H}$] & ion. state & Ref. & Tracing \\
& (km s$^{-1}$) & (cm$^{-3}$) & (cm$^{-3}$) & (cm$^{-2}$)  & & \\  \hline	
Low-velocity (LV) & -20 to 20 & - & - & 20.43 & neutral & 1 & Atomic/Molecular clouds in Orion \\
Intermediate-velocity (IV) & -80 to -20 & - & 0.16 & 18.44 & ionized & 1, 2, 3, 4 & Orion-Eridanus superbubble wall  \\
High-velocity (HV) & -120 to -80 & 3 $\times$ 10$^{-3}$ & 0.1 - 0.2 & 17.88  & ionized & 1, 2 & Nested shells (Barnard's Loop bubble, $\lambda$ Ori) \\ \hline \hline
\end{tabular}
\tablecomments{Listed quantities are (if known): observed velocity extents, $v_\m{LSR}$; pre-shock densities, $n_\m{0}$; post-shock densities, $n_\m{1}$; total hydrogen column densities, $N_\m{H}$; ionization state, and the components of the Orion-Eridanus superbubble that is traced by the specific gas velocities. The (column) densities quoted are observed in the line of sight towards $\zeta$ Ori. References are (1): \citet{welty_2002}, (2): \citet{cowie_1979}. (3): \citet{cowie_1978} (4): \citet{huang_1995}.}
\label{tab:connection}
\end{table*}

\subsubsection{Extent of the outer superbubble wall}\label{sec:extent}

The extent and connection of the IV gas to the superbubble wall (Sec. \ref{sec:absorption}) imply that the size of the Orion-Eridanus superbubble may be larger than previously thought. Indeed, Barnard's Loop is part of a separate closed bubble structure, suggesting that it is not associated with the large-scale expanding superbubble structure defined in earlier works  \citep[e.g.,][]{reynolds_1979,heiles_1999,bally_2008,pon_2014b}. Here, we argue that the outer H$\alpha$ filament towards the east in Fig. \ref{fig:superbubble} offers a prime candidate for representing the outer superbubble wall. 

We have concluded that Barnard's Loop is completely ionized by Orion OB1 and that ionizing photons are able to penetrate through. Moreover, in Secs. \ref{sec:bubble} and \ref{sec:specreg}, we have shown that Orion OB1 can provide the necessary ionizing photons to illuminate the H$\alpha$ emission seen in Fig. \ref{fig:superbubble}, indicating that the ionization front lies well outside the previously defined bounds of the Orion-Eridanus superbubble, which may be possible for superbubble walls in general \citep{basu_1999}. However, the large-scale, coherent H$\alpha$ filament to the north-east is indicative of a limb-brightened shell, rather than a mere haphazard ISM structure which happened to be illuminated by ionizing photons piercing through Barnard's Loop.

We note that often, the extent of superbubbles is traced through diffuse X-ray emission. The photo-electric cross section for 0.25 keV photons is $\sim$4 $\times$ 10$^{-21}$ cm$^{2}$ H$^{-1}$ \citep{morrison_1983}. By integrating the total foreground emission from the LAB survey \citep{kalberla_2005}, i.e., -400 $\textless$ $v_\m{LSR}$ $\textless$ 0 km s$^{-1}$ (which includes the near-side of the superbubble shell at velocities -40 $\textless$ $v_\m{LSR}$ $\textless$ 0 km s$^{-1}$; \citealt{brown_1995}), we infer that the 0.25 keV emission is confined to foreground \HI column densities of $N_\m{H}$ $\lesssim$ 3 $\times$ 10$^{19}$ cm$^{-2}$ (the superbubble cavity), where the optical depth for 0.25 keV X-ray photons is $\tau_\m{x}$ $\lesssim$ 0.1, while emission at 0.25 keV is absent from regions of $N_\m{H}$ $\gtrsim$ 2 $\times$ 10$^{20}$ cm$^{-2}$ (i.e, Barnard's Loop, and the east- and westside of the superbubble near the Galactic plane), where $\tau_\m{x}$ $\gtrsim$ 0.6. Thus, the clear anti-correlation of the 0.25 (and 0.75) keV emission (Fig. \ref{fig:superbubble}) with foreground HI limits the use of this tracer for many parts of the Orion-Eridanus region, especially towards the Galactic plane. 

The association of Barnard's Loop with a closed separate bubble, the IV gas that traces a region larger than the previously defined bounds of the Orion-Eridanus superbubble, and the mere presence of the large-scale, coherent H$\alpha$ filament to the north-east, has led us to believe that regions 1-5 (Fig. \ref{fig:apertures}) trace the actual outer shell of the Orion-Eridanus superbubble. Projected in the interior of the outer shell reside the Barnard's Loop bubble, the $\lambda$ Ori bubble, and the smaller-scale expanding bubbles discussed in Sec. \ref{sec:smallershells}. This indicates that the superbubble in its entirety consists of separate structures, or successive {\em nested shells}, possibly connected to a series of SN explosions originating from Orion OB1 over the past 15 Myr (\citealt{bally_2008}; Secs. \ref{sec:summaryextent} \& \ref{sec:scenario}). While hot plasma could exist between the H$\alpha$ filament and Barnard's loop, the detection of the accompanying X-ray emission is hampered because of foreground absorption. Our hypothesis sprouted from studying the energetics of gas and dust (Sec. \ref{sec:bubble} and Sec. \ref{sec:specreg}) and their morphological appearances (Fig. \ref{fig:superbubble}), combined with the connections between the different gas tracers with the components of the superbubble (Table \ref{tab:connection}). In the next subsection, we will discuss the temperature of the superbubble interior gas, as well as the global structure of the outer wall in order to substantiate our claim regarding the structure and extent of the Orion-Eridanus superbubble.

\subsubsection{Temperature of the superbubble interior and structure of the outer wall}\label{sec:tempint}

The evidence discussed in Sec. \ref{sec:extent} indicate that the outer superbubble wall lies outside of Barnard's Loop (Sec. \ref{sec:extent}), possibly extending towards the outer H$\alpha$ filament to the north-east (Fig. \ref{fig:superbubble}). Here, we quantify the temperature structure of the superbubble interior. First, we derive that the Barnard's Loop bubble is moving supersonically towards the east, from which we conclude that the temperature in this direction can not be as high as $\gtrsim$10$^6$ K. After comparison with the cooling timescale for a gas at $\textless$10$^6$ K, we infer that the medium between Barnard's Loop and the outer H$\alpha$ filament must have cooled to $\sim$10$^4$ K, appropriate for a gas photo-ionized by Orion OB1. We argue that mass-loading may cause temperature gradients within superbubble interiors, place our findings in the context of superbubble evolution, and discuss the structure of the outer wall.

The temperature structure of the superbubble interior can be estimated as follows. First, we define the isothermal sound speed, $c_\m{s}$ = ($kT$/$\mu$$m_{\m{H}}$)$^{1/2}$, where $\mu$ is the mean mass per hydrogen nucleus (= 0.61 for a fully ionized medium). Towards the west of the Barnard's Loop bubble, the temperature of the superbubble interior traced by the X-ray emitting gas is estimated at $T_\m{x}$ = 2.1 $\times$ 10$^6$ K \citep{burrows_1993}, giving $c_\m{s}$ = 170 km s$^{-1}$. We compare this with the Barnard's Loop bubble expanding at $\sim$100 km s$^{-1}$ (Sec. \ref{sec:blexp}), which implies that the remnant of the recent SN explosion (Secs. \ref{sec:blexp} \& \ref{sec:energetics}) is well within the radiative expansion phase (see below). During this phase, the velocity of the SNR decelerates with time as  $\propto$ $t^{-5/7}$ \citep{tielens_2005} if we assume that the ambient density in which the SNR propagates is distributed homogeneously. Adopting this time-velocity relation, the adiabatic expansion phase ($v_\m{exp}$ $\gtrsim$ 250 km s$^{-1}$) ended just below 10$^5$ yr, and the bright crescent of Barnard's Loop has been moving subsonically for $\sim$0.2 Myr (in case $c_\m{s}$ = 170 km s$^{-1}$). When moving subsonically, sound waves will travel ahead of the shell and the structure would dissolve and merge with the ISM within a fade-away timescale of \citep{mckee_1977, draine_2011b}

\begin{equation}
\label{eq:fadeaway}
\tau_\m{fade} = 1.87 E_\m{51}^{0.32} n^{-0.37} \left(\frac{c_\m{s}}{10 \m{\,km\, s^{-1}}}\right)^{-7/5} \m{\,Myr},
\end{equation} 	

where $E_\m{51}$ is the kinetic energy in numbers of 10$^{51}$ erg, and $n$ the ambient hydrogen number density. The current kinetic energy of the eastern half of the Barnard's Loop bubble measures $E_\m{51}$ $\sim$ 0.34 (Sec. \ref{sec:energetics}), part of which may already have been thermalized \citep{mac_low_1988}. However, the total kinetic energy delivered by a SN explosion typically does not exceed $E_\m{51}$ $\sim$ 1 \citep{veilleux_2005}, or $E_\m{51}$ $\sim$ 0.5 for the eastern half of the bubble. With $E_\m{51}$ = 0.5, and parameters for the X-ray emitting gas of the superbubble interior towards the west ($T_\m{x}$ = 2.1 $\times$ 10$^6$ K, $n_\m{x}$ = 6 $\times$ 10$^{-3}$ cm$^{-3}$; \citealt{burrows_1993}; Sec. \ref{sec:energetics}), we calculate $\tau_\m{fade}$ $\sim$ 0.2 Myr, similar to the time Barnard's Loop has been moving subsonically in case it is moving within hot gas at a temperature exceeding 10$^6$ K.

The above derived fadeaway timescale reveals that {\em if} Barnard's Loop would be moving to the east in an X-ray emitting gas such as that traced by {\em ROSAT} in the west part of the superbubble (Fig. \ref{fig:superbubble}), it must be long in the process of merging and dissolving with the hot gas. This actual process may cause the faintness of the H$\alpha$ emission of the Barnard's Loop bubble towards the west (Sec. \ref{sec:blexp}), where the superbubble interior temperature is known to be at a temperature of $T_\m{x}$ = 2.1 $\times$ 10$^6$ K (\citealt{burrows_1993}; Fig. \ref{fig:superbubble}). Clearly, this does not affect the eastern part of the Barnard's Loop bubble in the same way, as exemplified by the bright crescent of Barnard's Loop. Alternatively, there may be a density gradient within the superbubble interior that may cause the brightness difference of the Barnard's Loop bubble between its east and west side. Either way, the conclusion is that the Barnard's Loop bubble moves supersonically towards the east, in line with the findings of \citet{cowie_1979}, who argued that the HV gas (associated with the near-side of the Barnard's Loop bubble; Sec. \ref{sec:blexp}) traces a radiative shock. Consequently, the interior superbubble temperature towards the east can not be at $\gtrsim$10$^6$ K, as is observed for the X-ray emitting gas towards the west.

\begin{figure}
\centering
\includegraphics[width=9cm]{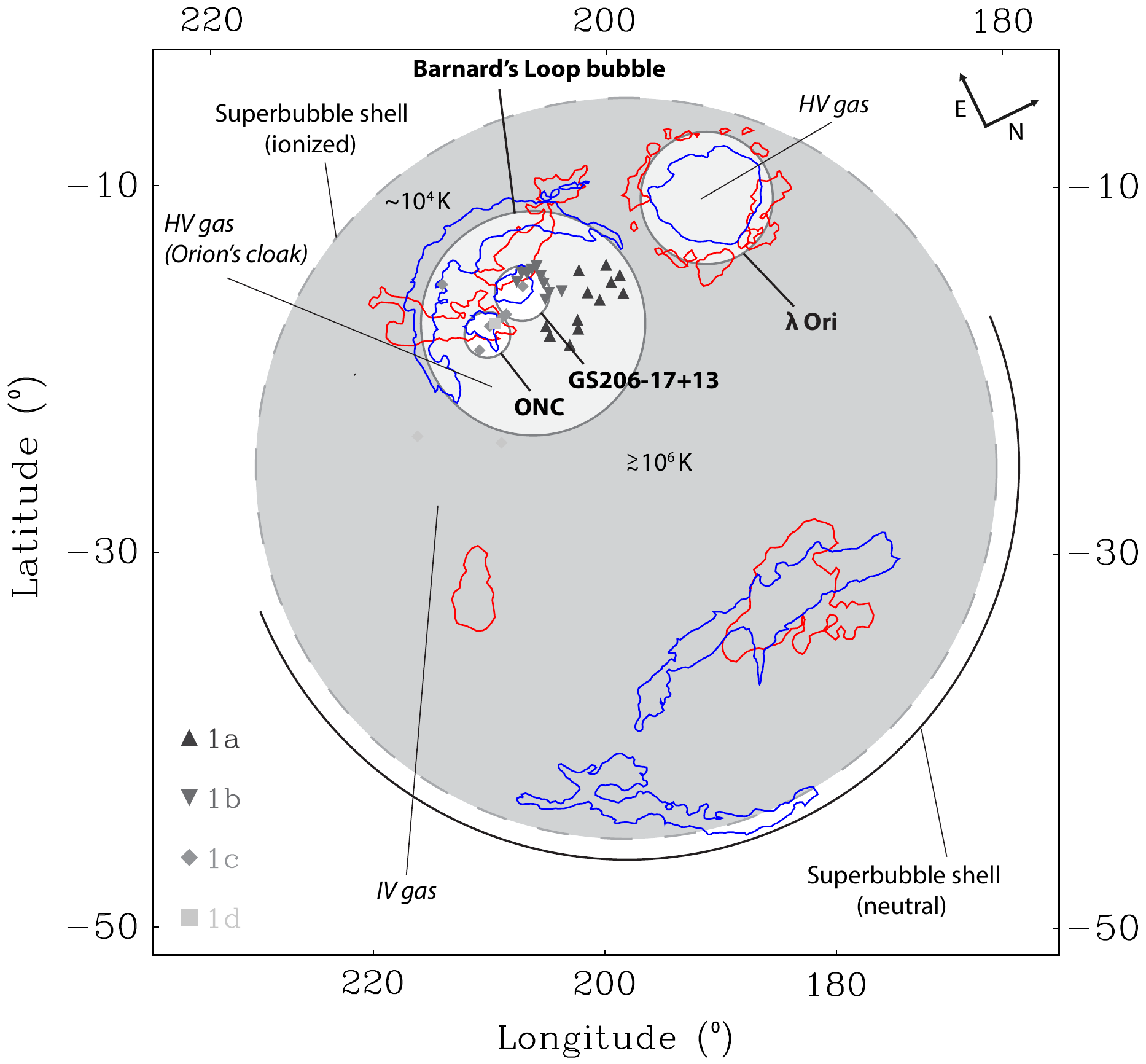} 
\caption{Schematic of the Orion-Eridanus superbubble and several of its major components. See Fig. \ref{fig:superbubble} for the nomenclatures of the shown structures. The proposed outer shell of the Orion-Eridanus superbubble (dashed circle), blown by a series of SNe from an old subpopulation of the Orion OB association, is traced by an intermediate-velocity (IV) shock that sweeps up the ambient ISM (gray area). Towards the west, the superbubble is surrounded by a shell of neutral ({\em solid black line}) swept-up material, whereas this shell is completely ionized  ({\em dashed gray line}) towards the east. The temperature in the superbubble interior seems to be lower towards the east ($\sim$10$^4$ K) in the direction of the Galactic plane, compared to the X-ray emitting gas in the west ($\gtrsim$10$^6$ K). The outer shell encompasses several `nested' smaller shells or bubbles ({\em solid gray circles}, denoted in boldface) that are more recently formed, such as the Barnard's loop bubble, traced by high-velocity (HV) gas. Other examples of bubbles that may have been triggered by ongoing activity in Orion OB1 are the $\lambda$ Ori bubble, GS206-17+13, and the shell surrounding the Orion Nebula cluster (ONC).} 
\label{fig:schematic}
\end{figure}	

Temperature differences within a superbubble interior can originate from thermal conduction and evaporation of the swept-up superbubble shell. This mechanism may cool the interior of the superbubble by increasing the interior density \citep{weaver_1977, mac_low_1988} such that radiative losses become important. Champagne flows and photo-ablation of molecular clouds enclosed in the superbubble can inject mass into the hot plasma, further lowering the temperature compared to that predicted by idealized models (a full discussion on the mass-loading mechanisms is deferred to Sec. \ref{sec:rejuvenation}). Gas below temperatures of 10$^6$ K cools rapidly. At an interior density of $n_\m{x}$ = 6 $\times$ 10$^{-3}$ cm$^{-3}$ \citep{burrows_1993}, the cooling timescale for a gas at 10$^5$ K equals $\tau_\m{cool}$ $\sim$ $kT/n\Lambda(T)$ $\sim$ 10$^5$ yr, where $\Lambda(T)$ $\sim$ 5 $\times$ 10$^{-22}$ erg cm$^{3}$ s$^{-1}$ is the average value of the cooling function for gas at temperatures 10$^4$ K $\lesssim$ $T$ $\lesssim$ 10$^6$ K \citep{dalgarno_1972}. Mass-loading of the interior will shorten the cooling timescale. The detection of \NV gas \citep{welty_2002} towards $\zeta$ Ori traces $\sim$10$^5$ K gas and confirms the presence of a thermally conductive layer \citep{chu_2008}, providing information on the temperature structure of the superbubble wall and the interior temperature, but it is unclear if this conductive layer traces the outer superbubble wall, or the interior shell associated with the Barnard's Loop bubble. Furthermore, it is uncertain if this layer extends towards the outer H$\alpha$ filament given the lack of coverage of absorption line studies in this region (Fig. \ref{fig:wham}). In any case, the cooling timescale is short compared to that of the estimated age of the superbubble ($\lesssim$ 8 Myr; \citealt{brown_1994}) and the supernova rate in Orion OB1 (1 - 1.5 Myr$^{-1}$; \citealt{bally_2008}; Sec. \ref{sec:scenario}) that may re-heat the interior gas (see below and Sec. 
\ref{sec:rejuvenation}). Consequently, if the temperature drops below 10$^6$ K, the medium between Barnard's Loop and the outer filament will settle at a temperature of $\sim$10$^4$ K, as it is kept photo-ionized by Orion OB1 by photons piercing through Barnard's Loop (Sec. \ref{sec:specreg}). Depending on the internal density, the medium between Barnard's Loop and the outer filament may not be in pressure equilibrium with the (higher-density) outer H$\alpha$ filament. 

In the early stages of its evolution, an outer superbubble wall is defined by a shell of swept-up material, the expansion of which is driven by an overpressurized interior medium \citep{tielens_2005}. The superbubble initially expands through the Sedov-Taylor solution, but as the interior medium expands adiabatically, its pressure will decrease and the expansion slows down. The radiative phase commences when the expansion velocity decreases to $\sim$250 km s$^{-1}$ or a post-shock temperature of $\sim$10$^6$ K \citep{tielens_2005}. At this point, radiative losses will become important, and the interior temperature (and pressure) will further drop. When the interior pressure reaches that of the ambient medium, the expansion of the superbubble wall will coast at constant radial momentum, thus slowing down as it sweeps up material from its surroundings. When the expansion velocity reaches the local sound speed, the superbubble wall will disappear and merge with the ISM within a fadeaway timescale (Eq. \ref{eq:fadeaway}). The interior pressure may even drop below that of the ambient medium if the cooling timescale ($\sim$10$^5$ yr) is shorter than the pressure timescale or sound-crossing time. Currently, the distance between Barnard's Loop and the outer Halpha filament is roughly $\sim$100 pc, such that the pressure timescale can be anywhere up to $\sim$10 Myr for a 10$^4$ K medium, depending on the exact moment the region cooled towards this temperature. However, the medium between Barnard's Loop and the outer filament must have cooled within the time between successive expanding shells (regulated by the SN rate; 1 - 1.5 Myr$^{-1}$; \citealt{bally_2008}). Thus, it may well be that the pressure inside the interior cannot adjust rapidly enough to the change in temperature and, in case no mass-loading occurs, it can reach a lower pressure compared to the ambient galactic medium. In case of an underpressurized interior, the expansion of the superbubble wall may halt and even reverse its movement if the pressure difference between the interior and the ambient medium is strong enough.
 
Given that the medium between Barnard's Loop and the outer H$\alpha$ filament is at 10$^4$ K and we are not able to constrain its density, one may question if this medium in fact represents the WIM or WNM. In this case, the outer H$\alpha$ filament is not the superbubble wall, but merely an ISM feature illuminated by photons escaping from Barnard's Loop. However, we have revealed that Barnard's Loop is part of a separate closed structure (Sec. \ref{sec:blexp}), and argued that the superbubble wall extends beyond Barnard's Loop (addressed in Secs. \ref{sec:absorption} \& \ref{sec:extent}). Surely, if one accepts that Barnard's Loop is expanding at 100 km s$^{-1}$ and represents a SNR of age 0.3 Myr (Secs. \ref{sec:blexp} \& \ref{sec:energetics}), it may not be surprising that the superbubble wall lies further outwards as measured from Orion OB1, given that star formation has occurred over 15 Myr in Orion and that 10-20 SNe have gone of in its past \citep{bally_2008}. Here, we argue that successive SN explosions may lead to the appearance of successive shells. In this scenario, the large-scale, coherent H$\alpha$ filament to the north-east represents a prime candidate to trace the actual outer superbubble wall. We defer a further discussion on the idea of successive shells and the evolution of the outer superbubble wall to Sec. \ref{sec:scenario}.  

We note that the outer H$\alpha$ filament (Fig. \ref{fig:superbubble}) does not contain a neutral shell that is expected for a sweeping superbubble wall (Fig. \ref{fig:cuts} \& \ref{fig:apertures}). The lack of a neutral shell may simply suggest that the swept-up ISM did not contain enough material to fully absorb the ionizing photons from Orion OB1. Alternatively, the east side of the superbubble, which is racing up the density ramp towards the Galactic plane, may not be moving supersonically anymore and has changed character from a shock wave to a sound wave, rendering it unable to sweep up additional mass. Using Eq. \ref{eq:fadeaway} and $E_\m{51}$ $\sim$ 10 (the total kinetic energy exterted Orion OB1; \citealt{brown_1995,bally_2008}), we calculate for the east side of the bubble expanding towards the WNM of the Galactic plane ($n$ $\sim$ 0.5 cm$^{-3}$, $c_\m{s}$ $\sim$ 10 km s$^{-1}$), $\tau_\m{fade}$ $\simeq$ 4 Myr, well below the estimated age of the superbubble ($\lesssim$ 8 Myr; \citealt{brown_1994}). Thus, we may be observing the east side of the superbubble as it is in the process of dissolving, already having lost its neutral shell and currently being completely ionized before it merges with the local material of the ISM. Eq. \ref{eq:fadeaway} does not apply to the west side of the superbubble as its expansion velocity (traced through \HI; $\sim$40 km s$^{-1}$; \citealt{brown_1995}) implies that it moves supersonically through the WNM/WIM. However, for the ionized east side of the superbubble the expansion velocity is not constrained, and the question whether or not it is moving supersonically remains unanswered.

\subsubsection{Summary: the structure of the Orion-Eridanus superbubble}\label{sec:summaryextent}

We now have obtained the complete set of tools to derive an updated picture of the Orion-Eridanus region which we show in Fig. \ref{fig:schematic}. Our results demonstrate that the superbubble consists of separate structures or {\em nested shells}, superimposed along line of sight. In particular, we have recognized a separate bubble (amongst others; Sec. \ref{sec:shells}), evident as the HV gas and the Barnard's Loop bubble in H$\alpha$ (Fig \ref{fig:wham}). This bubble expands {\em inside} the pre-existing cavity of the Orion-Eridanus superbubble, given its pre-shock density, its embedded nature in the Orion-Eridanus region (Fig. \ref{fig:schematic}), and its likely origin through feedback from Orion OB1. The realization that Barnard's Loop is not part of the superbubble wall (Sec. \ref{sec:blexp}), the IV gas that seems to trace a region larger than the previously defined bounds of the Orion-Eridanus superbubble (Sec. \ref{sec:absorption}), and the presence of the large-scale, coherent H$\alpha$ filament to the north-east, has led us to believe that the extent of the Orion-Eridanus superbubble is larger than previously throught. We have argued that the temperature of the superbubble interior between Barnard's Loop and the outer H$\alpha$ filament to the north-east can not be as high as that observed for the X-ray emitting gas towards the west \citep{burrows_1993} and, therefore, must have settled at 10$^4$ K (Sec. \ref{sec:tempint}). The expansion of this outer H$\alpha$ filament may not be pressure-driven anymore. While this is an inevitable outcome during late stages of superbubbles and SNRs \citep{tielens_2005}, the interior pressure drop may be accelerated through the effects of mass loading and radiative energy losses (Sec. \ref{sec:rejuvenation}). The observations (Fig. \ref{fig:superbubble}) indicate that the outer superbubble wall is surrounded by a shell of \HI towards the west, whereas it may be completely ionized towards the east (Sec. \ref{sec:tempint}).

We stress that the association of gas velocities with components in the superbubble (Table \ref{tab:connection}) are largely based on the analysis of a dozen of sightlines throughout the Orion side of the superbubble (Fig. \ref{fig:wham}). It is not clear how much mass is associated with these velocities (Sec. \ref{sec:energetics}). Moreover, some of the gas tracers are not seen in particular lines of sight. This may be because we are tracing foreground stars, such as the sightline towards $\gamma$ Ori located at 77 pc \citep{van_leeuwen_2007}, which could provide a constraint to the elongation of the bubble towards the Local Bubble \citep[e.g.,][]{bally_2008,pon_2014b}, or incomplete spectral coverage \citep{cowie_1979}. Alternatively, the distribution of the gas components might be patchy \citep{cowie_1979}. Surely, the connections laid out in Table \ref{tab:connection} may be oversimplified, as the Orion region is highly complex, containing dozens of filamentary clouds, and must bear the marks of multiple SNRs \citep{bally_2008} that are breaking out of a region that contains a lot of substructure. Further observations are needed to firmly constrain the extent of the distinct gas components in order to derive the detailed structure of the Orion-Eridanus superbubble, and to recognize the existence of conductive layers that would provide more insight in the temperature distribution throughout the superbubble interior. 

The observations of the gas and dust have revealed that Orion OB1 can be the source of ionization of the H$\alpha$ filaments encompassed within the entire region shown in Fig. \ref{fig:superbubble} (Sec. \ref{sec:apertures}). We have connected the IV gas with the outer superbubble wall. We note that the pre-shock density of the IV is not known. However, given a post-shock density of 0.16 cm$^{-3}$ \citep{welty_2002} and shock velocity of 40 km s$^{-1}$ \citep{brown_1995}, a shock moving through the WNM ($T$ = 8000 K, $c_\m{s}$ $\sim$ 10 km$^{-1}$) would have a pre-shock density of order $\sim$0.01 cm$^{-3}$ (J-type shocks increase densities by $\mathcal{M}^2$; \citealt{tielens_2005}), which is somewhat low for the WNM in the Galactic plane. Higher densities for the Orion-Eridanus shell of 1 - 5 cm$^{-3}$ are quoted towards the Eridanus filaments in the west \citep[e.g.,][]{reynolds_1979}, indicating that the expansion occurs somewhat anisotropically, which can also be inferred by appreciating the complex morphology of the superbubble. In any case, the shock velocity of the IV gas ($\sim$ 40 km s$^{-1}$) is not fast enough to ionize hydrogen: therefore, the Eridanus filaments (and the IV gas) are photo-ionized by Orion OB1, which was already noted by \citet{reynolds_1979} and \citet{welty_2002}. We note that \citet{pon_2014b} explored other mechanisms of ionization, but concluded that Orion OB1 most likely causes the observed ionization state of the IV gas. When the spherical geometry from Fig. \ref{fig:schematic} is adopted for the outer superbubble wall, and assuming a diameter of the supershell of 45 degrees (equivalent to a radius of 160 pc) with expansion velocity of $\sim$ 40 km s$^{-1}$, the dynamical age of the shell is $\sim$ 4 Myr, consistent with the age of the Orion OB1a subgroup, minus a main sequence lifetime of 3 - 5 Myr for its most massive stars \citep{bally_2008}. Naturally, stellar winds during the first few Myr of the lifetime of the OB association may have contributed to the formation and expansion of the superbubble wall as well.

In conclusion, considering that star formation had been occurring for 10 - 15 Myr in Orion OB1 and the presence of at least four distinct subgroups \citep{blaauw_1964,bally_2008}, it may not be surprising that the Orion-Eridanus superbubble consists of a set of nested shells. Our results connect well with recent work from \citet{schlafly_2015}, who show that the Orion molecular clouds are in turn part of an ancient ring of dust. Possibly, the ring represents the remainder of a large bubble that pushed the Orion molecular clouds out of the Galactic plane towards their current location. Still, the hierarchy of bubbles does not end at the possible progenitor of the Orion-Eridanus superbubble defined by \citet{schlafly_2015}. Indeed, it has been argued that the Gould's Belt that include all nearby OB associations are associated with a large expanding ancient (30 - 60 Myr) supershell known as Lindblad's ring (see \citealt{bally_2008}, and references therein). The Gould's Belt of stars and OB associations may therefore represent sequential star formation when the Lindblad's ring cooled and collapsed to form dark clouds out of which the evolution of the Orion region commenced.

\subsection{The nested shells and their relation to Orion OB1}\label{sec:relationob1}

The recent discovery that the Orion molecular clouds may be part of an ancient ring of dust \citep{schlafly_2015}, the uncovering of an extended foreground population towards Orion A \citep{alves_2012,bouy_2014}, and the mysterious origin of some of the brightest stars in Orion ($\alpha$ Ori, $\beta$ Ori, and $\kappa$ Ori; \citealt{bally_2008}) illustrates that our current understanding of the Orion region is still limited. \citet{bally_2008} proposed that the bright stars may be part of a foreground group about 150 pc in front of Orion OB1, and together constitute the three most massive members of this population still present to date. Alternatively, some of the massive stars from Orion OB1 may have been ejected as runaways (space velocities in excess of $\textgreater$ 40 km s$^{-1}$) in the direction of the Local Bubble before exploding as SNe: up to 25\% of the OB stars end up as runaways \citep{zinnecker_2007}. Such stars could be linked to the odd geometry of the Orion-Eridanus superbubble \citep{pon_2014b}, which appears to be elongated towards the Local Bubble.

One must keep in mind that most of the distance determinations towards the Orion stellar populations are based on photometric indicators \citep[e.g.][]{brown_1994}. Although the Hipparcos mission determined parallaxes for many stars in the Orion region, the distance to the young populations ($\sim$ 300 - 400 pc) translates to Hipparcos parallax errors of 30 - 40\% (given a precision of 1 milliarcsecond; \citealt{perryman_1997}). Thus the detailed distribution of the stellar population along the line of sight can, as of yet, not be determined \citep[see also][]{dezeeuw_1999}. In this respect, the Gaia mission \citep{debruijne_2012,perryman_2001} is expected to dramatically improve the distance and age determinations of the stars and subgroups in Orion. The entire Orion region lies well within the range where Gaia parallaxes will be accurate to 10\% or better over a large range of spectral types. This will allow to constrain the relationship between the stellar content and the nested shells seen in the Orion-Eridanus superbubble, and to address the importance of triggered star formation \citep{elmegreen_1977} in Orion, for which one needs {\em precise} measurements of proper motion, radial velocities, and ages.

\subsection{Evolution of the Orion-Eridanus superbubble}\label{sec:rejuvenation}

In this section, we will exploit the results drawn from the previous sections to derive an updated picture of the evolution of the Orion-Eridanus superbubble, regulated by a balance between processes that mass-load the superbubble interior, controlling its temperature and density structure \citep{cowie_1981,hartquist_1986}, and feedback from Orion OB1 that sweep up and collect the injected interior mass in expanding shells before thermalizing or cooling \citep{mac_low_1988}. These processes determine the momentum of the superbubble wall and the thermal pressure of the interior, which drives the superbubble expansion out into the ambient Galactic medium.

First, in Sec. \ref{sec:energetics} we summarize masses and energetics as observed from different components of the Orion-Eridanus superbubble. From this, we conclude that the Barnard's Loop bubble is a SNR (associated with a SN explosion that occurred 3 $\times$ 10$^5$ yr ago; Sec. \ref{sec:hvivgas}; \citealt{cowie_1979}), and argue in Sec. \ref{sec:scenario} that the SNR has swept through the inner parts of the superbubble to collect mass that has been introduced through champagne flows and thermally evaporating clouds. Behind the Barnard's Loop bubble, champagne flows from IC 434, the Orion Nebula, and $\lambda$ Ori continue to mass-load the superbubble interior until a next SN explosion will start the cycle all over again. The net effect is a gradual disruption of the star-forming reservoir and the subsequent outward transportation of processed material, where it will ultimately pressurize the superbubble interior, or add momentum to the outer superbubble wall. We will argue that at the current rate, the OMCs will be able to power the expansion and evolution of the Orion-Eridanus superbubble for another 20 - 30 Myrs, before they are depleted and run out of steam, at which point the superbubble will disappear and merge with the surrounding ISM.

\subsubsection{Masses and energetics of the Orion-Eridanus superbubble}\label{sec:energetics}

\begin{table*}
\centering
\caption{Masses and energetics of the Orion-Eridanus superbubble}
\begin{tabular}{l|c|c|c|c|l|l}\hline \hline
& $M$ & $v_\m{exp}$ & $E_\m{th}$ & $E_\m{kin}$ & Form & Reference \\
& (10$^3$ $M_{\odot}$) & (km s$^{-1}$) & (10$^{49}$ erg) & (10$^{49}$ erg) &  &  \\  \hline	
\multicolumn{6}{l}{{\em Hot gas ($\gtrsim$10$^6$ K)}} \\ \hline
Superbubble interior & 1.8 & - & 62 & - & Thermal energy & \citet{burrows_1993}  \\
Barnard's Loop bubble interior & (0.02) & - & (0.65) & - & Thermal energy & This work \\ \hline
\multicolumn{6}{l}{{\em Ionized gas (10$^4$ K)}} \\ \hline
Superbubble & 84 & 15$^{(1)}$ & 14 & 19 & Swept-up shell & \citet{reynolds_1979} \\ 
Barnard's Loop bubble & 6.7 & 100 & 1 & 67 & Swept-up shell & This work \\ 
$\lambda$ Ori & 2 - 6 & 30 & 0.5 & 3 & Champagne flow & \citet{reich_1978,van_buren_1986} \\
GS206-17+13 & 3 & 30 & 0.5 & 3 & Champagne flow & This work \\
IC 434 & 0.1 & 30 & 0.02 & 0.1 & Champagne flow & \citet{ochsendorf_2015} \\ 
Orion Nebula & 0.02 & 30 & 0.003 & 0.02 & Champagne flow & \citet{wilson_1997} \\ \hline
\multicolumn{6}{l}{{\em Neutral gas (10$^2$ K)}} \\ \hline
Superbubble & 250 & 40 & - & 370 & Swept-up shell & \citet{brown_1995} \\ 
GS206-17+13 & 3.4 & 8 & - & 0.2$^{(2)}$ & Swept-up shell & \citet{ochsendorf_2015} \\ 
Orion Nebula (Veil) & 2.3 & 2 & - & 0.01 & Swept-up shell & \citet{vanderwerf_2012}; This work \\ \hline
\multicolumn{6}{l}{{\em Molecular gas (10 K)}} \\ \hline
OMC A & 105 & -  & - & - & Molecular cloud & \citet{wilson_2005} \\ 
OMC B & 82 & - & - & - & Molecular cloud & \citet{wilson_2005} \\ 
$\lambda$ Ori & 11 & 16.5 & - & 6 & Swept-up shell & \citet{lang_1998} \\ \hline 
\multicolumn{6}{l}{{\em Stars}} \\ \hline
Ori 1a & 1.6 & -  & - & - & Stellar cluster & \citet{brown_1994} \\ 
Ori 1b & 1.3 & - & - & - & Stellar cluster & \citet{brown_1994} \\ 
Ori 1c & 1.8 & -  & - & - & Stellar cluster & \citet{brown_1994} \\ 
Ori 1d (ONC) & 1.8 & - & - & - & Stellar cluster & \citet{hillenbrand_1997} \\ 
Ori 1d (NGC2024) & 0.2 & - & - & - & Stellar cluster & \citet{comeron_1996} \\ 
$\sigma$ Ori & 0.2 & - & - & - & Stellar cluster & \citet{sherry_2004} \\
($\lambda$ Ori) & (0.6) & -  & - & - & Stellar cluster & \citet{barrado_2004} \\ \hline \hline
\end{tabular}
\tablecomments{Listed quantities are: mass, $M$; the expansion velocity, $v_\m{exp}$; and the thermal ($E_\m{th}$) and kinetic ($E_\m{kin}$) energy. $^{(1)}$:  the expansion velocity in H$\alpha$ and \HI are not measured at the same position causing the discrepancy between both numbers. $^{(2)}$: This value has been revised downwards compared to \citet{ochsendorf_2015}, who determined the kinetic energy assuming an abrupt energy input appropriate for supernova remnants \citep{chevalier_1974}. Here, we argue that the energy input has been continuous through stellar winds (see text) and instead use $E_\m{kin}$ = 1/2$Mv_\m{exp}^2$.}
\label{tab:energetics}
\end{table*}

{\em Hot gas} ($\gtrsim$ 10$^6$ K; Table \ref{tab:energetics}) from the superbubble interior has been characterized by \citet{burrows_1993}, who inferred a temperature of $T_\m{x}$ $\sim$ 2 $\times$ 10$^{6}$ K and a total luminosity at X-ray wavelengths of $L_\m{x}$ $\approx$ 8.0 $\times$ 10$^{35}$ erg s$^{-1}$, when assumed that the X-ray emitting region is spherical and its actual `hook-shaped' appearance results after foreground absorption by MBM 18 and MBM 20 (Fig. \ref{fig:superbubble}; \citealt{burrows_1993}). The authors deduce the following parameters for the hot gas: when 400 pc is taken as the distance towards the superbubble, the radius of the X-ray emitting region is $r_\m{x}$ = 70 pc, its density is $n_\m{x}$ = 6 $\times$ 10$^{-3}$ cm$^{-3}$ \citep{burrows_1993}, and contains a total mass $M_\m{x}$ = 230 M$_\m{\odot}$ after adopting a spherical geometry. However, if we adopt the (larger) geometry from Fig. \ref{fig:schematic}, and account for possible missing X-ray emission because of foreground absorption (Fig. \ref{fig:superbubble} and Sec. \ref{sec:extent}), the total mass within a region of $r_\m{x}$ = 160 pc would amount to $M_\m{x}$ = 1800 M$_\m{\odot}$. Table \ref{tab:energetics} also denotes a value for the hot gas inside the Barnard's Loop bubble of radius 35 pc in case it were filled with gas of the same characteristics as observed in the superbubble. As noted before (Sec. \ref{sec:extent}), the high column densities for the Barnard's Loop region prohibits the detection of diffuse X-rays towards the east side of the Barnard's Loop bubble. Hot gas is detected towards the west side where the \HI foreground column is much lower, but it is unclear if this emission traces hot gas from the Barnard's Loop bubble or the entire Orion-Eridanus superbubble.

{\em Ionized gas} ($\sim$$10^4$ K; Table \ref{tab:energetics}) in the Orion-Eridanus region is associated with the outer H$\alpha$ filaments of the superbubble, the Barnard's Loop bubble shell, and the champagne flows of the $\lambda$ Ori bubble, GS206-17+13, the IC 434 emission nebula, and the Orion Nebula (Sec. \ref{sec:apertures} \& Sec. \ref{sec:shells}). For Barnard's Loop, the total ionized mass is estimated at $M$ = 6700 M$_\m{\odot}$ by adopting a half-sphere geometry with thickness 1$^\circ$, outer radius of 7$^\circ$, and density $n_\m{H}$ = 3.4 cm$^{-3}$ (Sec. \ref{sec:bl}). The western half of the complete bubble (Fig. \ref{fig:wham}) is much fainter, which implies that the density in the Barnard's Loop bubble varies significantly between both parts (Sec. \ref{sec:tempint}). Assuming $v_\m{exp}$ = 100 km s$^{-1}$ (Sec. \ref{sec:shells}), $E_\m{kin}$ =  3.4 $\times$ 10$^{50}$ erg. The total energy that has been injected in the region can then be estimated at $E_\m{kin}$ =  6.7 $\times$ 10$^{50}$ erg by assuming spherical symmetry. This energy should be compared to the typical kinetic energy provided per SN (10$^{50}$ - 10$^{51}$ erg, depending on the efficiency at which it couples to the ISM; \citealt{veilleux_2005}) as well as the wind mechanical energy input. By adopting typical wind parameters ($\dot{M}$ = 1 $\times$ 10$^{-6}$ M$_\odot$ yr$^{-1}$, $v_\m{\infty}$ = 2000 km s$^{-1}$) for the 7 most-massive stars in the Orion OB association, and assuming the lifetime of Orion's Cloak at 3 $\times$ 10$^5$ yr (Sec. \ref{sec:hvivgas}; \citealt{cowie_1979}), the total mechanical luminosity exerted by the Orion OB association is $L_\m{mech}$ = 8.4 $\times$ 10$^{49}$ erg, of which $\sim$20\% couples to the surrounding medium as kinetic energy \citep{van_buren_1986}. Thus, the (combined efforts of) stellar winds from the Orion OB1 stars seem insufficient to provide the measured kinetic energy of the Barnard's Loop bubble and we associate the Barnard's Loop bubble with a recent supernova explosion.

Table \ref{tab:energetics} reflects the omnipresence of champagne flows in the Orion region, consistent with the findings of \citet{ochsendorf_2014b}, who showed that  \HII region champagne flows are ubiquitous throughout the Galactic plane, and illustrated the importance of this phase to the turbulent structure of the ISM. Still, the amount of material that is eroded from the molecular clouds and delivered by champagne flows in the form of kinetic energy of the fast moving ionized gas is difficult to determine, and estimates over the lifetime of the star will depend most critically on the ionizing luminosity and main sequence lifetime of the ionizing star \citep[e.g.,][]{whitworth_1979}. Here, we will follow \citet{tielens_2005} and adopt a `typical' blister phase that will accelerate $M_\m{\odot}$ $\sim$ 3 $\times$ 10$^3$ M$_\m{\odot}$ of ionized gas to 30 km s$^{-1}$, providing a kinetic energy of $E_\m{kin}$ $\sim$ 3 $\times$ 10$^{49}$ erg. We use this value for the champagne flow of the GS206-17+13 bubble that has occurred in the past (Sec. \ref{sec:smallershells}). For comparison, the ionized interior density of the $\lambda$ Ori region, that seems to be on the verge of breaking out (Sec. \ref{sec:lambdaori}), is quoted to be between 2000 M$_\odot$ \citep{van_buren_1986} and 6000 M$_\m{\odot}$ \citep{reich_1978}: this would add a similar amount of mass and kinetic energy to the superbubble interior as the GS206-17+13 shell did. The Orion nebula has just recently entered the champagne phase (Sec. \ref{sec:smallershells}) and only contains some $\sim$ 20 M$_\m{\odot}$ of ionized gas \citep{wilson_1997}. The IC 434 region is a special case, as the champagne flow originates from a chance encounter between the $\sigma$ Ori star cluster and the Orion B molecular cloud \citep{ochsendorf_2014a,ochsendorf_2015}. Still, the star has already eroded 100 $M_\m{\odot}$ and will add another 350 $M_\m{\odot}$ to the ionized gas before it plunges into the molecular cloud (for a current distance of 3.5 pc and a closing velocity of $\sim$10 km s$^{-1}$; \citealt{ochsendorf_2015}).

{\em Neutral gas} ($\sim$10$^2$ K; Table \ref{tab:energetics}) is present in the superbubble wall \citep{brown_1995}, and the slowly expanding shells of the $\lambda$ Ori region \citep{lang_1998}, the GS206-17+13 bubble \citep{ochsendorf_2015}, and the Orion Nebula that is encapsulated by the neutral `Veil' \citep[e.g.,][]{vanderwerf_2012}. The GS206-17+13 is a stellar wind-blown-bubble (Sec. \ref{sec:smallershells}). The Orion nebula Veil contains a total neutral column density of $N_\m{H}$ = 5.8 $\times$ 10$^{21}$ cm$^{-2}$,  has a size of at least 1.5 pc, as it is common to the Orion Nebula and M43, and shows only minor blue shifted velocities with respect to the background OMC ($\sim$2 km s$^{-1}$). This implies that the mechanical feedback of the Trapezium through, e.g., the champagne flow or stellar winds has not yet coupled as kinetic energy to the surrounding neutral gas, consistent with the young age of the Trapezium stars ($\textless$1 Myr; \citealt{hillenbrand_1997}) and, perhaps, the very small dynamical time inferred for the Orion Nebula since the onset of the champagne flow ($\sim$1.5$\times$ 10$^4$ yr; \citealt{o'dell_2009}). Here, we estimate the total \HI mass for the Orion Veil by assuming that the measured column density by \citet{vanderwerf_2012} is homogeneously distributed over a shell with a radius of 2 pc, large enough to cover both the M43 region as well as the entire Extended Orion Nebula \citep{guedel_2008}.

{\em Molecular gas} ($\sim$$10$ K; Table \ref{tab:energetics}) is detected in dozens of small cometary clouds, but largely concentrates in the Orion molecular clouds, which constitute the main reservoir of the dense gas out of which star formation occurs. As described by, e.g., \citet{bally_2008}, the ages and locations of the subgroups in Orion OB1 indicate that star formation has propagated through the `proto-Orion cloud' sequentially. The kinetic energy associated with the expansion of the $\lambda$ Ori molecular shell (6 $\times$ 10$^{49}$ erg) is consistent with expansion driven by either the overpressure of photo-ionized gas \citep{spitzer_1978} that deposits some 10$^{49}$ - 10$^{50}$ erg to the surrounding medium \citep{ochsendorf_2014b} over the lifetime of $\lambda$ Ori ($\sim$10 Myr), or a SN explosion \citep{veilleux_2005}, or a wind-blown origin \citep{weaver_1977}, as $L_\m{mech}$ = 2 $\times$ 10$^{50}$ erg for $\lambda$ Ori \citep{van_buren_1986} assuming an age of 5 Myr \citep{bally_2008}, a fraction of which will couple as kinetic energy. Some previous works favor a SN origin for the $\lambda$ Ori bubble, which would also explain the current space motion of the $\lambda$ Ori star (see Sec. \ref{sec:lambdaori}). 
 
{\em Stars} are listed for comparison, where we have included the separate subgroups from the Orion OB1 association (1a, 1b, 1c, and 1d; \citealt{blaauw_1964, brown_1994}). The masses in the 1a, 1b, and 1c subgroups have been determined using the results in \citet{brown_1994}, who constructed the initial mass function (IMF) and through this reported the number of stars initially found in the 2 - 120 M$_{\odot}$ mass range for each subgroup. We complement this by estimating the contribution from low mass stars below the completion limit of the \citet{brown_1994} study (0.08 - 2 M$_{\odot}$) using a standard IMF \citep{kroupa_2001}, normalized to the results from \citet{brown_1994}. Table \ref{tab:energetics} lists the resulting total masses, assuming an upper stellar mass limit set by the age of each subgroup. In Orion OB1a, stars more massive than 13 M$_\odot$ have exploded as SNe, while the (younger) age of the Orion 1b and 1c subgroups imply an upper mass limit of about 20 M$_\odot$ \citep{bally_2008}. Orion OB1d is the youngest subgroup and includes many star-forming clusters \citep{bally_2008}, and here we only include the two largest components, the Orion Nebular Cluster in OMC A, and NGC 2024 in OMC B. The $\sigma$ Ori group is noteworthy here, as it has traditionally been assigned to Orion OB1b based on its spatial location amongst the Belt stars, but its traversed space motion reveals that the cluster might have originated from the Orion OB1c region \citep{ochsendorf_2015}. We have bracketed the $\lambda$ Ori region in Table \ref{tab:energetics} as it is separated from the star clusters that are associated with the OMCs, and does not directly influence the scenario that will be outlined in Sec. \ref{sec:scenario}.
 
{\em Summary:} Most of the mass of the Orion-Eridanus superbubble resides in the outer superbubble wall, while the nested shells that process and sweep-up the interior material (see Sec. \ref{sec:scenario}) constitute only a small part. The molecular clouds still represent a large reservoir of mass, comparable to that located in the superbubble wall. Most of the energy is found in the form of kinetic energy of the expansion of the superbubble wall. Still, about 10 - 15\% of the total energy is located in thermal energy of the hot gas of the superbubble interior, and a comparable amount lies within the kinetic energy of the ionized gas. The thermal energy of the hot gas created by fast shocks is large compared to the thermal energy of the 10$^4$ K ionized gas delivered through photo-ionization from Orion OB1. Finally, when all subgroups in Table \ref{tab:energetics} are combined, stars comprise some $\sim$6.9 $\times$ 10$^3$ M$_\m{\odot}$ or 3 - 4\% of the molecular material within the OMCs.

\subsubsection{Mass-loading and cleansing of the interior}\label{sec:scenario}

The expanding shells in the interior of the Orion-Eridanus superbubble are the manifestation of feedback from Orion OB1. When moving supersonically, the shockwaves associated with the shells sweep up the interior mass of the superbubble, acting as a mechanism to remove any material that may have been introduced to the medium in the past. In this respect, the mass (6.7 $\times$ 10$^3$ M$_\odot$) and radius (35 pc) of the shell surrounding the Barnard's Loop bubble can be combined to show that the average density of the swept-up medium contained $\sim$1 cm$^{-3}$. We note that the density of the swept up material only slightly exceeds what would be expected for the WNM of the Galaxy, which would be consistent with the Barnard's Loop bubble being a superbubble wall expanding into the ambient Galactic medium. However, we render this scenario unlikely given the short dynamical timescale of the Barnard's Loop bubble ($\sim$0.3 Myr; Sec. \ref{sec:blexp}) compared to the Orion star forming region ($\sim$15 Myr; \citealt{bally_2008}), the pre-shock density of the HV gas ($\sim$ 3 $\times$ 10$^{-3}$ cm$^{-3}$; \citealt{welty_2002}) and its nested appearance within the entire Orion-Eridanus superbubble (Figs. \ref{fig:superbubble} \& \ref{fig:wham}). The increased density is about two orders of magnitude above that what is expected for a superbubble interior ($\sim$10$^{-2}$ cm$^{-3}$; \citealt{weaver_1977}), which provides convincing evidence that the superbubble interior must have been mass-loaded since the last supernova passed through and cleansed the interior.

Photo-ionization feedback from Orion OB1 ultimately creates champagne flows that channel 10$^4$ K gas from the \HII regions into the superbubble cavity, where it will mix with the hot gas ($\gtrsim$10$^{6}$; \citealt{chu_2008}). The injected mass will significantly lower the temperature of the superbubble interior and increase its density. The total mass delivered to the interior by a typical champagne flow over the lifetime of a massive star ($\sim$3 $\times$ 10$^3$ M$_\m{\odot}$) only slightly exceeds the mass that makes up the hot gas of the entire superbubble (1.8 $\times$ 10$^3$ M$_\m{\odot}$). However, this extra mass will be injected in a small portion of the interior (see below). In addition, champagne flows will dump kinetic energy in the form of turbulent motion into the hot cavity. In principle, the injected turbulent energy is more than the thermal energy of the hot gas provided by the SN that lead to the formation of the Barnard's Loop bubble (Table \ref{tab:energetics}). However, dissipation of this turbulence will lead to temperatures of only $\sim$ 10$^4$ K (for a flow speed of $\sim$ 30 km s$^{-1}$; Table \ref{tab:energetics}; \citealt{tielens_2005}). The cooling function peaks at these temperatures \citep{dalgarno_1972} such that this excess energy will be mainly radiated away. Nevertheless, adding and mixing in the champagne gas will mass-load and cool the hot superbubble interior gas. In addition, thermal evaporation from the dozens of clouds that are submersed in the hot gas of the superbubble, most notably the Orion Molecular clouds, will contribute to the mass-loading of the hot gas. It is difficult to constrain the relative contributions to the mass-loading of the hot gas through thermal and photo-evaporation, as the thermal evaporation rate is a steep function on temperature ($\propto$ $T^{5/2}$; \citealt{cowie_1977}), while photo-evaporation rates depend on the ionizing luminosity and main sequence lifetime of the ionizing star \citep[e.g.,][]{whitworth_1979, bodenheimer_1979}. Nonetheless, we can infer that the {\em combined} effects of the mass-loading mechanisms have introduced some 6.7 $\times$ 10$^3$ M$_\m{\odot}$ of material into the superbubble cavity over the past 3 $\times$ 10$^5$ yr given by the mass and age of the H$\alpha$ shell associated with the Barnard's Loop bubble that acts to cleanse the interior of the superbubble close to the OMCs. These numbers imply an average evaporation rate (or cloud destruction rate) of $\sim$1 $\times$ 10$^{-2}$ M$_\m{\odot}$ yr$^{-1}$, which is an order of magnitude above the current photo-erosion rate in the Orion Nebula, if we use a total ionized gas mass of 20 $M_\m{\odot}$ \citep{wilson_1997} and a dynamical lifetime of 1.5 $\times$ 10$^4$ yr \citep{o'dell_2009}. Part of this discrepancy may be because the Orion nebula still resides in an early phase of evolution and the region has not fully broken out of its natal cloud. Nonetheless, mass-loading of the superbubble interior increased the density by a factor of $\sim$100 compared to that expected in case the region within the Barnard's Loop bubble were filled with hot gas of the same characteristics as observed in the superbubble (Table \ref{tab:energetics}). Over time, this will have a rigorous effect on the thermal behavior of the interior gas.

We note that the pre-shock density derived for the HV gas associated with the Barnard's Loop bubble in the line of sight towards $\zeta$ Ori (Sec. \ref{sec:shells}) is 3 $\times$ 10$^{-3}$ cm$^{-3}$, while we have derived an average density of 1 cm$^{-3}$ for the medium in which the Barnard's Loop bubble has expanded in its 3 $\times$ 10$^5$ yr lifetime. Surely, the mass-loading mechanisms described here concentrate towards the centre of star formation, i.e., the OMCs. There are believed to have been 10 - 20 SN explosions within Orion in the last 12 Myr \citep{bally_2008}. With an expansion velocity of $\sim$30 km s$^{-1}$ for the gas introduced through photoablation and champagne flows (Table \ref{tab:energetics}), a supernova rate of 1 - 1.5 Myr$^{-1}$ will not allow the injected mass to disperse and fill the entire volume of the superbubble before being swept-up by a subsequent SNR. Thus, the increased densities due to mass-loading from the OMCs will be mainly concentrated within 30 - 45 pc surrounding the OB association (consistent with the radius of the Barnard's Loop bubble), whereas the surrounding medium will have a much lower density. The low temperature in the eastern part of the superbubble between Barnard's Loop and region 1 (Sec. \ref{sec:tempint}) may also be caused by the effects of mass-loading. This would imply that other `routes' of mass-loading affect the density and temperature structure of this region, such as enclosed clouds toward the Galactic plane, the straight evaporation of the superbubble wall, or SNRs that dissipate before reaching the superbubble wall (see below). However, as we have not been able to constrain the density structure of this region, the importance of mass-loading remains unclear.

In contrast to the shell of the Barnard's Loop bubble that is kept photo-ionized by Orion OB1 (Sec. \ref{sec:specreg}), the neutral shells surrounding $\lambda$ Ori and GS206-17+13 have had the chance to cool radiatively and collapse to form dense media. While they may add to the mass-loading of the superbubble interior through thermal evaporation, their slow expansion renders them unimportant compared to the total energetics of the superbubble (Table \ref{tab:energetics}). Still, these dense shells may be the site for a new generation of stars as they are prone to gravitational instabilities \citep{zavagno_2007,pomares_2009,martins_2010}. 

The supernova remnant associated with the Barnard's Loop bubble has exploded $\sim$3 $\times$ 10$^{5}$ yr ago and currently expands at 100 km s$^{-1}$, implying that it is well in the momentum-conserving snow plow phase \citep[e.g.,][]{tielens_2005}. The post-shock gas contains densities ($n$ = 3.4 cm$^{-3}$; Sec. \ref{sec:bl}) and temperatures ($T$ = 6000 K; \citealt{heiles_2000,o'dell_2011,madsen_2006}) that would allow the gas to cool rapidly ($\tau_\m{cool}$ $\sim$ $kT/n\Lambda(T)$ $\sim$ 4 $\times$ 10$^4$ yr, where $\Lambda(T)$ $\approx$ 6 $\times$ 10$^{-25}$ erg cm$^{3}$ s$^{-1}$; \citealt{dalgarno_1972}). However, Barnard's Loop is kept photo-ionized by Orion OB1 (Sec. \ref{sec:specreg}), prohibiting the cooling and collapse of the gas to form a dense neutral shell. In this case, the shockwave will continue to gather mass until it becomes subsonic, from which point it will dissipate and pressurize the interior, thus effectively displacing material from the origin of the blast wave towards the outskirts of the superbubble interior. 

To get a feeling how far the displacement of material will reach, we can compare radii and expansion velocities of the Barnard's Loop bubble ($r$ = 35 pc, $v_\m{exp}$ = 100 km s$^{-1}$) and the superbubble ($r$ = 160 pc, $v_\m{exp}$ = 40 km s$^{-1}$) to estimate that the Barnard's Loop bubble will catch up with the outer wall within $\sim$2 Myr if both structures decelerate at the same rate (in the radiative expansion phase, the velocity of the SNR decelerates with time as  $\propto$ $t^{-5/7}$; \citealt{tielens_2005}). At this moment, the isothermal sound speed in the outer regions of the superbubble towards the east is $\sim$10 km s$^{-1}$ (for $T$ $\sim$ 10$^4$ K; Sec. \ref{sec:tempint}). If one assumes the lifetime of Barnard's Loop at 3 $\times$ 10$^5$ yr, and adopt the time-velocity relation of the SNR above, the adiabatic expansion phase ($v_\m{exp}$ $\gtrsim$ 250 km s$^{-1}$) ended just below 10$^5$ yr and it will take another $\sim$7 Myr to reach 10 km s$^{-1}$ if the density ahead of the SNR is homeogeneous. This timescale far exceeds that derived for the Barnard's Loop bubble to catch up with the superbubble wall. Thus, Barnard's Loop will be able to cross the superbubble radius and reach the outer wall, where the shockwave will eventually thermalize and the swept-up material will transfer its mass and momentum by condensing onto the superbubble wall \citep{mac_low_1988}. The outer superbubble wall may therefore be accelerated as various interior shells such as Barnard's loop catch up to it. In between such pulses, it could be coasting and slowing down as it sweeps up surrounding gas in a momentum conserving interaction, or it may be dissolving in case it moves subsonically (Sec. \ref{sec:tempint}). 

We note that densities in the outer parts of the superbubble are smaller compared to the mass-loaded region close to the OB assocation ($\sim$ 1 cm$^{-3}$; see above). In this case, the resulting pressure drop will accelerate the interior SNR once again when it reaches the edge of the mass-loading region, which already seems to be the case for the Barnard's Loop bubble, given the pre-shock density of 3 $\times$ 10$^{-3}$ cm$^{-3}$ in the line of sight towards $\zeta$ Ori \citep{welty_2002}. Indeed, towards the west side of the superbubble, the temperature of the interior is much higher and the SNR will accelerate, but it may go subsonic and dissolve earlier compared to the east side of the bubble: this may be at the root of the faintness of the H$\alpha$ emitting shell in this direction (Sec. \ref{sec:tempint}). Towards the east, the density gradient could be smaller because of the lower temperature that may be accompanied by an increase in density, and the amount of acceleration will decrease accordingly.

If 10 - 20 SN explosions occurred in the Orion region in the past 12 Myr \citep{bally_2008}, and each SNR adds $\sim$7 $\times$ 10$^3$ M$_\odot$ of material to the superbubble wall (Table \ref{tab:energetics}), this will amount to 0.7 - 1.4 $\times$ 10$^5$ M$_\odot$ and account for a significant fraction (20 - 40 \%) of its total mass. Thus, the mass of the superbubble wall without the addition from SNRs would account for 1.9 - 2.6 $\times$ 10$^5$ $M_\m{\odot}$. With a radius of 160 pc, we can estimate the average ISM density into which the superbubble has expanded at $n$ $\sim$ 0.15 - 0.2 cm$^{-3}$, consistent with the WNM within one or two scale heights of the Galactic plane. Finally, as the supernova rate is of order for the Barnard's Loop bubble to catch up with the outer wall, it may not be necessary to expect multiple layers of expanding shells between the Orion OB association and the superbubble wall.

\subsubsection{Efficiency of star formation in the Orion molecular clouds}

Given a total mass of $\sim$2 $\times$ 10$^{5}$ M$_\m{\odot}$ for the OMCs combined, a supernova rate of 1 - 1.5 Myr$^{-1}$ \citep{bally_2008} and swept-up mass of $\sim$7 $\times$ 10$^3$ M$_\odot$ per SNR (the mass of the H$\alpha$ emitting part of the Barnard's Loop bubble), we estimate that at this rate the reservoir of cloud material will be depleted within 20 - 30 Myr through thermal evaporation and erosion from champagne flows alone, similar to calculated cloud lifetimes from \citet{williams_1997}. Thus, material is effectively removed from the molecular clouds and incorporated in expanding shells that might form stars of their own. Furthermore, the molecular clouds will be disrupted by each supernova blast wave that mark the end of a cycle of star formation. All of these effects should be considered if one attempts to simulate the lifecycle of molecular clouds and the efficiency of star formation, which is well known to be limited by various forms of stellar feedback (for a recent review, see \citealt{krumholz_2014}).

The gradual erosion of the OMCs puts a firm upper limit on the lifetime of the molecular clouds. Over the past 10 - 15 Myr, some $\sim$10$^5$ M$_\m{\odot}$ has already been lost from the OMCs to the superbubble wall, which implies that the original molecular reservoir (the `proto-Orion' cloud) contained some $\sim$3 $\times$ 10$^5$ M$_\m{\odot}$ (Table \ref{tab:energetics}). At present, the major subgroups of the Orion OB association \citep{blaauw_1964, brown_1994} constitute some 6.9 $\times$ 10$^3$ M$_\m{\odot}$ of stellar mass. The ages of the subgroups imply that many stars have already exploded as supernova (Sec. \ref{sec:rejuvenation}; \citealt{bally_2008}), and we estimate from the adopted IMF that about 20\% of the initial mass of the 1a, 1b, and 1c subgroups has been lost through SN explosions, increasing the total mass that has been incorporated into stars over the last 10 - 15 Myr towards $\sim$ 7.4 $\times$ 10$^3$ M$_\m{\odot}$. With a remaining cloud lifetime of the OMCs of 20 - 30 Myr, at the current rate, some $\sim$2.2 $\times$ 10$^4$ M$_\m{\odot}$ or 7.5\% of the molecular mass can be converted into stars over the lifetime of the clouds consistent with predictions by, e.g., \citet{williams_1997}. However, it is unclear if star formation will proceed similarly as stellar feedback may trigger star formation {\em and} disrupt the clouds, while the molecular reservoir steadily diminishes (see, e.g., \citealt{evans_2009,hopkins_2011,krumholz_2014} for more thorough discussions regarding the mechanisms that involve star formation rates and molecular cloud lifetimes).

\subsubsection{Summary: the rejuvenation of the Orion-Eridanus superbubble}

We have argued that appreciable amounts of mass will be removed from the molecular clouds and be mass-loaded into the interior of the superbubble through destructive champagne flows and thermal evaporation of clouds embedded in the hot gas of the superbubble interior. Explosive feedback from stellar winds and, in particular, supernovae take care of the subsequent transportation of the mass-loaded material towards the outskirts of the superbubble through the formation of expanding nested shells. For the Barnard's Loop bubble towards the east, photo-ionization by Orion OB1 does not allow it to radiatively cool and collapse, and it will plaster the mass and add momentum to the superbubble wall that will continue to drive the expansion of the superbubble. In case it would go subsonic before reaching the outer wall, the shell would dissolve, thermalize, and pressurize the interior. This cycle repeats as long as there is ongoing star formation and OB stars can maintain (the photo-ionization of) the cleansing shells that deposit the mass towards the outer regions. 

\subsection{Dust processing in superbubbles}

An interesting consequence of the constant mass-loading of the superbubble is that each supernova explosion within the superbubble cavity will interact with a significant amount of `fresh' interstellar gas that carries along dust in its wake through entrainment. The efficiency of dust entrainment in photo-evaporation flows was investigated by \citet{ochsendorf_2015} for the IC 434 champagne flow, revealing that the dust-to-gas ratio within the flow is similar to that seen in the diffuse ISM ($\sim$0.01). Thus, each single SNR from the Orion OB1 association will encounter a new reservoir of gas and dust within the superbubble cavity, introduced by champagne flows from the \HII regions and from photo-evaporation flows of the thermally evaporating clouds. This finding contrasts with that of \citet{mckee_1989}, who previously estimated the timescale for SNR shocks to destroy a typical dust grain within the ISM. \citet{mckee_1989} assumed that only the first supernova within an OB association would destroy an amount of dust equivalent to that located in an ISM volume containing $\sim$1300 M$_\m{\odot}$ of gas. Following supernovae from the OB association would hammer surrounding gas already cleansed of its dust component. Following \citet{mckee_1989}, the {\em effective} SN rate is equal to

\begin{equation}
\label{eq:heating}
\tau_\m{SN}^{-1} = \big(q_\m{I}f_\m{I}  + \left[f_\m{II,OB}q_\m{II,OB} + f_\m{II,field}q_\m{II,field}\right]f_\m{II} \big)\kappa_\m{SN} \m{\,\,\,yr}.
\end{equation} 

Here, $q_\m{I}$ is the correlation factor of the SNe of type Ia ($\sim$0.38; \citealt{heiles_1987}), relating the location of the SN to their position above the disk of the Galaxy (i.e., it measures the effective interaction of the SNe with the amount of gas in their environment), and $f_\m{I}$ is the fraction of supernovae of type Ia ($\sim$0.5; \citealt{narayan_1987}). Similarly, $f_\m{II}$  (= 1 - $f_\m{I}$) measures the fraction of supernova of type II, part of which occur in OB associations, described by $f_\m{II,OB}$ ($\sim$ 0.75; see \citealt{zinnecker_2007}, and references therein), and in the field, denoted with $f_\m{II,field}$ (= 1 - $f_\m{II,OB}$), originating from runaways and isolated stars with correlation factor $q_\m{II,field}$ ($\sim$0.6; using values for the scale height of the WNM by \citealt{heiles_1987}). Finally, $\kappa_\m{SN}$ is the intrinsic SN rate in the Galaxy ($\sim$2 $\times$ 10$^{-2}$ yr$^{-1}$; \citealt{diehl_2006}).

Under the assumption that superbubbles do not replenish their dust content after the first SNR, \citet{mckee_1989} arrived at a correlation factor of $q_\m{II,OB}$ = 0.1 and, together with a rather low value for the fraction of type II SNe in OB associations, $f_\m{II,OB}$ = 0.5, derived an effective SN rate of $\tau_\m{SN}$ = 125 yr. In this work, we have shown that dust entrainment accompanying mass-loading of the superbubble interior replenishes the dust content inside the superbubble after each SNR from the OB association, i.e., $q_\m{II,OB}$ = 1. In addition, we use the more recently derived fraction of OB stars that are located in OB associations from \citet{zinnecker_2007}, $f_\m{II,OB}$ = 0.75. These changes lead to shorter effective supernova rate of $\tau_\m{SN}$ = 90 yr, implying that the SNRs within superbubbles destroy dust more efficiently and, subsequently, the dust lifetime against SNR shocks must be revised downwards by 30\%.

The processing of dust inside the superbubble can also be probed directly from depletion studies. \citet{welty_2002} noted that in the HV and IV gas towards $\zeta$ Ori, the gas-phase abundances of Al, Si and Fe are slightly elevated compared to that observed in warm, diffuse clouds. However, the carbon abundance is {\em significantly} raised in the HV gas to near-solar abundances. For comparison, depletion studies of the Cold Neutral Medium (CNM) and the WNM show a pattern where a thin coating on interstellar dust gets sputtered and re-accreted, as the grains cycle from clouds to the diffuse phase of the ISM \citep{tielens_1998,tielens_2013}. Whilst in the WNM, the reservoir of gas and dust typically encounters a 100 km s$^{-1}$ passing shock, destroying some 10 - 30\% of the silicate volume \citep{jones_1996}. However, these shock velocities would only destroy 15\% of the carbonaceous dust and lead to minor variations in the carbon depletion, as $\sim$ 50\% of the carbon is in the gas phase \citep{cardelli_1996,sofia_2004}. Therefore, instead of passing shocks, we attribute the enhanced carbon abundance to sputtering of the grains inside the hot gas of the Orion-Eridanus superbubble. The sputtering timescale $\tau_\m{sput}$ \citep{tielens_1994} in the X-ray emitting gas ($T$ $\sim$ 10$^6$ K, $n$ $\sim$ 0.01 cm$^{-3}$) implies that large 0.1 $\mu$m silicate grains are hardly affected ($\tau_\m{sput}$ $\textgreater$ 20 Myr), yet, small $\sim$ 10 $\AA$ (carbonaceous) grains are quickly eroded ($\tau_\m{sput}$ $\textless$ 1 Myr), releasing the carbon into the gas-phase. We argue that this may be a general characteristic for dust that is processed inside superbubbles. Indeed, dust entering the heliosphere also shows a significant increase of gas-phase carbon, implying that the grains are sputtered within the hot gas of the Local Bubble \citep{frisch_2013}.

\section{Summary}\label{sec:summary}

The general picture that emerges from the multitude of data is that the Orion-Eridanus superbubble is larger and more complex than previously thought, consisting of a series of nested shells which we depict in Fig. \ref{fig:schematic}. With the data currently at hand, we have discussed the following structures, ordered in size from bottom up:

\begin{enumerate}
\item[-] The youngest shell surrounds the Orion Nebula cluster located in Orion A (age $\textless$ 1 Myr; \citealt{bally_2008}).
\item[-] Next, there is the 2 to 4 degree HI and dust shell GS206-17+13, approximately centered on the $\sigma$ Ori cluster, but likely blown by stellar winds from the Belt stars (age $\sim$ 5 Myr; \citealt{ochsendorf_2015}). 
\item[-] Barnard's Loop is part of a complete bubble structure not related to the outer edge of the Orion-Eridanus superbubble. It is instead associated with a supernova remnant (age $\sim$ 3 $\times$ 10$^5$ yr), and is connected with high-velocity gas (`Orion's Cloak') detected in absorption studies \citep{cowie_1979,welty_2002}. All of the previous shells are projected in the interior of this young 7 degree radius bubble, which expands at high velocity ($v_\m{exp}$ $\sim$ 100 km s$^{-1}$). 
\item[-] The Barnard's Loop bubble and the $\lambda$ Ori bubble ($\sim$ 5 Myr; \citealt{bally_2008}) expand in the interior of the $\sim$ 45 $\times$ 45 degree shell that is related to intermediate-velocity gas from the aforementioned absorption studies and created by the collective effects of Orion's stars: the Orion-Eridanus superbubble (age $\sim$ 5 - 10 Myr).
\end{enumerate}

As we can recognize several distinct subgroups and well defined clusters, star formation in the Orion-Eridanus region must have been highly episodic over the last 10 - 15 Myr. It may therefore not be surprising that the Orion-Eridanus superbubble consists of a set of nested shells. Each `burst' of star formation may have produced a subgroup consisting of several clusters that disrupted parts of the pre-existing molecular cloud. The formation of each subgroup and the resulting UV radiation can ionize large amounts from the dense reservoir of molecular gas in the region that, together with thermal evaporation of the molecular clouds, constantly mass-load and cool the hot gas inside the superbubble cavity. Stellar winds and supernovae accelerate, sweep-up, and compress these `poisoned' plasmas in an episodic fashion to form nested shells within the Orion-Eridanus superbubble, such as the Barnard's Loop bubble, the $\lambda$ Ori bubble, the GS206-17+13 shell, and the Orion nebula (see Fig. \ref{fig:schematic}). The shells may cool, collapse, fragment, and be incorporated in a next generation of stars. However, for the Barnard's Loop bubble, this shell-cooling is inhibited by photo-ionization from the OB association. We have noted that in its turn, the Orion-Eridanus superbubble has previously been associated with the ancient supershell known as Lindblad's ring, which may have cooled and collapsed to form the Gould's Belt of stars that includes the Orion OB association, extending the hierarchy of nested shells and bubbles towards greater sizes and into history. The continuous replenishment of dust in the superbubble cavity through entrainment in ionized flows has lead us to conclude that dust processing from interior supernova remnants is more efficient than previously thought. To conclude, the cycle of mass-loading, cleansing, and star formation ceases when feedback has disrupted the molecular reservoir, from which the superbubble will disappear and merge with the ISM. 

\begin{acknowledgements} 
{\bf Acknowledgements. }The authors would like to thank the anonymous referee for detailed comments that significantly increased the quality of this paper. Studies of interstellar dust and chemistry at Leiden Observatory are supported through advanced ERC grant 246976 from the European Research Council, through a grant by the Dutch Science Agency, NWO, as part of the Dutch Astrochemistry Network, and through the Spinoza premie from the Dutch Science Agency, NWO.
\end{acknowledgements}

% for the bibliography, at the end
%\bibliographystyle{aa} % style aa.bst
%\bibliography{orion_eridanus.bib} % your references Yourfile.bib

\end{document}